\documentclass[12pt]{article}
\usepackage{arxiv}
\usepackage[utf8]{inputenc} 
\usepackage[english]{babel}
\usepackage{url}            
\usepackage{booktabs}       
\usepackage{tabularx}
\usepackage{amsfonts}       
\usepackage{nicefrac}       
\usepackage{microtype}      
\usepackage{graphicx}
\usepackage{natbib}
\setcitestyle{aysep={}}
\usepackage{soul}
\usepackage{siunitx}
\usepackage{threeparttable}
\usepackage{adjustbox}
\usepackage{multirow}
\usepackage{float}

\usepackage{subfig}
\usepackage{graphics}
\usepackage{subcaption} 
\usepackage{amsmath}
\usepackage{amssymb}
\usepackage{enumitem}
\usepackage[table]{xcolor}
\usepackage{color}
\usepackage{amsthm}
\usepackage{dsfont}
\usepackage{lscape}
\usepackage{appendix}
\usepackage{dirtytalk}
\usepackage{afterpage}
\usepackage{comment}
\usepackage{setspace}
\usepackage{bm}
\usepackage{array}
\usepackage[section]{placeins}
\usepackage[hyperfootnotes=false]{hyperref}       
\usepackage{doi}

\usepackage{subfig}
\def\vx{\textbf{x}}

\def\mu{\mathbf{u}}

\def\Exp{\text{E}}

\usepackage{bm}

\allowdisplaybreaks

\newcolumntype{L}[1]{>{\raggedright\let\newline\\\arraybackslash\hspace{0pt}}m{#1}}
\newcolumntype{C}[1]{>{\centering\let\newline\\\arraybackslash\hspace{0pt}}m{#1}}
\newcolumntype{R}[1]{>{\raggedleft\let\newline\\\arraybackslash\hspace{0pt}}m{#1}}
\usepackage{MnSymbol}

\usepackage{xcolor}
\definecolor{aqua}{rgb}{0.0, 1.0, 1.0}
\definecolor{babyblue}{rgb}{0.54, 0.81, 0.94} 
\definecolor{ballblue}{rgb}{0.13, 0.67, 0.8}
\definecolor{bondiblue}{rgb}{0.0, 0.58, 0.71}
\definecolor{mycorrection}{rgb}{255,0,206}
\definecolor{mygray}{RGB}{192,192,192}					
\definecolor{mygreen}{RGB}{44,85,17}					
\definecolor{myblue}{RGB}{36,68,92}					
\definecolor{myweb}{RGB}{233,233,192}					
\definecolor{mybrown}{RGB}{194,164,113}				
\definecolor{myred}{RGB}{255,66,56}	
\usepackage{hyperref}

\hypersetup{
    colorlinks=true,
    linkcolor=blue,
    filecolor=black,      
    urlcolor=cyan,
    citecolor =blue,
}

\newcolumntype{m}{>{\hsize=.6\hsize}X}
\newcolumntype{s}{>{\hsize=.4\hsize}X}

\title{The effects of Temperature and Rainfall Anomalies on Macroeconomic Variables: The case of Mexico}

\author{ 
    Lenin Arango-Castillo\thanks{Corresponding author.}\\
  \texttt{l.arango.castillo@protonmail.com} \\
       \And
  Francisco J. Martinez-Ramirez \\
  \texttt{franciscomartnez54@gmail.com} \\
}

\usepackage{cleveref}
\begin{document}
\maketitle

\onehalfspacing
\clearpage
\newpage

\begin{abstract}
This paper measures the effects of temperature and precipitation anomalies on Mexican headline inflation and total GDP per capita using two different approaches. We use data from all states in Mexico aggregated in seven geographical regions. We estimate the effects on inflation using Panel Local Projections method, while the effects on GDP per capita, using the Panel Autoregressive Distributed Lag Model. Our results indicate that neither temperature or precipitation anomalies have a statistically significant effect on GDP per capita, headline inflation, or their components.
\end{abstract}

\doublespacing
\cleardoublepage 
\pagenumbering{arabic}
\vspace{-1cm}

\section{Introduction}
\label{sec:intro}

Until very recently, governments and policymakers had remained skeptical about the impacts of climate change. However, the frequency, intensity, and duration of extreme weather-related events (e.g., heat waves, heavy precipitation, droughts, tropical cyclones, etc.) have increased, and this has led the international scientific community to investigate what is causing them and what are their consequences. While these phenomena can be determined, for example, by changes in the Earth's orbit \citep{Andersson2020}, there is now a consensus that the adoption of new technologies based on fossil fuels (e.g. coal, oil, and natural gas) since the Industrial Revolution has induced an increasing release of greenhouse gases (GHG) into the atmosphere, which has led to global warming and, hence, to extreme weather-related events (\citealt{Andersson2020} and \citealt{bundesbank2022}). 

In this paper, we use two different approaches to estimate the effect of extreme weather-related events on Mexico's macroeconomic variables, specifically total GDP per capita and the production of three sectors (primary, secondary, and tertiary), as well as headline inflation, three sub-indices of core inflation (services, non-food items, and food, beverages, and tobacco), and two sub-indices of non-core inflation (agriculture and energy). We estimate the effects on GDP per capita using \cite{kahn2021long}'s methodology based on the Autoregressive Distributed Lag Model, while we estimate the effects on inflation using \cite{jorda2005estimation}'s Local Projections method. The GDP per capita effects are estimated considering the positive and negative values of the weather-related variables as in \cite{kahn2021long} and \cite{liu2025macroeconomic}.

The effects of climate change are important for a country like Mexico, as it is one of the most biodiverse countries in the world, containing between 10 and 12 percent of the world's global species, and one of the most populous, with a population slightly above 128 million people. This unique combination of vast natural resources and a large population has allowed Mexico, according to the OECD, to profit from the exploitation of natural resources such as forests, soil, water, and fisheries, which have become important sectors in Mexico's economic development; as the country is especially endowed with abundant energy resources, both fossil and renewable, and is still a net exporter of crude oil. As a result, there have been multiple attempts to model the consequences of climate change on the Mexican economy, with most of them focusing on the effects on the primary sector (see \cite{ibarran2025} and the references therein). Most of these studies focus on estimating the present value of the total costs producers will face due to output losses caused by extreme weather events, or under conditions of recurring temperature increases.

Some of these studies in the literature estimate that \say{climate change} will have serious effects on Mexican GDP, for instance, \cite{estrada2022impacts} estimated that by the end of the 21st century, the present value of losses in some crops, including but not limited to maize, rice, sorghum, and wheat will reach twice the current total agricultural production value. Another set of studies focuses on the effects of climate change on total or sectorial growth.\footnote{\,See for example, \cite{mendelsohn2010ricardian}; \cite{arellano2023temperature}, \cite{arellano2023weather_}} For example, \cite{arellano2025temperature} estimate that economic growth, measured as the growth rate of total GDP, might be reduced by 0.4 percentage points, on average, under an intermediate scenario of climate change.\footnote{\,In \cite{arellano2025temperature}, temperature and precipitation variables are Mexico-specific, not global variables. Moreover, although the authors constructed temperature and precipitation normals, their model seemingly relies on levels rather than anomalies \citep[Eq. (1) on page 12]{arellano2025temperature}.} However, the majority of the literature assessed demonstrates pervasive methodological shortcomings that call into question the robustness of the reported results \citep[Appendix A.2]{kahn2021long}. To the best of our knowledge, no previous study for Mexico has used local temperature and precipitation anomalies to examine their effects on both GDP per capita and inflation, nor applied the methodological combination employed in this work. Using local climate variables in a subnational panel, we study how local temperature and precipitation fluctuations are associated with changes in two Mexican macroeconomic variables.

The majority of the existing empirical literature has focused on examining the impact of weather-related events on economic growth and productivity. These studies have been conducted for different regions and countries, and, more recently, for sectors other than agriculture, a sector highly exposed to outdoor weather conditions \citep{Colacito2019} and whose share in a country's Gross Domestic Product (GDP) is generally small (\citealt{dell2012}; \citealt{sudarshan2014economic}; \citealt{zhang2015temperature}; \citealt{zhang2018temperature}).\footnote{\,For example, agriculture accounts for 1~percent of the United States (US)'s GDP and for 10 percent of China's GDP, while the manufacturing sector accounts for 12 percent and 32 percent of the US's and China's GDP, respectively \citep{zhang2018temperature}.} The main results have shown, for example, that an increase of 1 degree Celsius, depending on the study and country under analysis, does reduce economic growth, productivity, investment, and industrial production, mostly in poor or less developed countries (see \citealt{dell2012}; \citealt{letta2019weather}; \citealt{acevedo2020effects}; \citealt{Andersson2020}; and \citealt{alvi2021assessing}). The effect of higher temperatures on advanced economies seems to be less discernible (\citealt{dell2012}; \citealt{Colacito2019}).

The impact of weather-related events on inflation has received comparatively less attention (\citealt{parker2018impact}; \citealt{faccia2021feeling}; \citealt{kotz2023impact}; \citealt{ehlers2025}). However, examining this linkage is of utmost importance given the dual pressures of climate change and the climate policies designed to mitigate it, both of which could affect central banks' ability to control inflation (\citealt{batten2018climate}; \citealt{Andersson2020}; \citealt{bundesbank2022}; \citealt{kotz2023impact}). On the one hand, global warming can lead to extreme weather-related events that may destroy crops, buildings, and infrastructure, causing a temporary shortage of goods and services and, therefore, higher inflation (\citealt{heinen2019price}; \citealt{dafermos2021price}; \citealt{faccia2021feeling}). In addition, higher temperatures and more frequent extreme weather-related events can negatively affect productivity and dampen long-term aggregate potential output growth. This, in turn, can reduce equilibrium real interest rates and diminish the ``room for manoeuvre for conventional monetary policy measures'' (\citealt{faccia2021feeling}; \citealt{bundesbank2022}; \citealt{cevik2023eye}). From the demand side, rehabilitation and reconstruction following such events can also lead to higher prices \citep{dafermos2021price}. On the other hand, climate policies aimed at mitigating global warming, such as the implementation of emissions trading schemes or taxes on high-carbon activities, may also exert upward pressure on prices (\citealt{faccia2021feeling}; \citealt{BNPParibas2022}; \citealt{bundesbank2022}).

Downward pressures on prices can also emerge in the aftermath of these extreme weather-related events. The destruction of assets after these events may reduce households' and firms' wealth and, therefore, their consumption and investment (
\citealt{dafermos2021price};  \citealt{faccia2021feeling}; \citealt{bundesbank2022};  and \citealt{cevik2023eye}). This effect on prices can occur even if assets are insured against natural disasters. The reasons for this are twofold: higher insurance costs related to more frequent extreme weather-related events, and a lower credit supply by banks as a result of higher loan defaults following a disaster, which may further reduce agents' wealth and, therefore, their consumption and investment (\citealt{batten2018climate}; \citealt{parker2018impact}; \citealt {dafermos2021price}; \citealt{bundesbank2022}; and \citealt{cevik2023eye}).  

Overall, it can be seen that climate change and the extreme weather-related events that have emerged do affect the central bank's ability to maintain price stability and meet its inflation target. This, in turn, implies that climate change also has distributional consequences: economic agents' purchasing power will be affected, but those employed in sectors most affected by climate change are the ones that will experience the largest negative effect on their income and wealth (\citealt{dafermos2021price} and \citealt{kotz2023impact}). Hence, assessments on the effects of climate change on prices seem crucial to shed some light on which economic sectors are the most vulnerable to climate risks, on the magnitude and duration of the effects, and on policies that could help mitigate such effects.  



Our contribution to the literature is twofold. First, we estimate the effects of extreme weather-related events on GDP per capita and inflation.  Second, we use anomalies of the weather-related variables, instead of the levels and squares of these variables as in several papers \citep[see Appendix~A.2]{kahn2021long}. This is important since variables such as temperature exhibit a trend and their inclusion in levels (and squares) in a specification could introduce a trend in the outcome variable that did not even exist, leading to biased results. 

Our results indicate that neither temperature nor precipitation anomalies have a statistically significant effect on GDP per capita, headline inflation, or their components. These findings are consistent with existing evidence in the literature that found no effect or a negligible one on inflation or GDP per capita. For example, \citet[Graph 3]{ehlers2025} shows that the average worldwide annual impact of climate-related disasters—linked to temperature anomalies or storms—amounts to less than 0.1\% of GDP for Mexico. Moreover, in a set of Latin American countries including Mexico, \citet[Graph 4]{ehlers2025} finds that storms have no statistically significant effect on GDP growth. However, storms do have a statistically significant effect on energy inflation one or two months after occurrence, but no significant effect on food inflation \citet[Graph 5]{ehlers2025}. Under the Representative Concentration Pathways (RCP) 2.6 and 8.5,\footnote{\, The projected persistent increases in global temperatures differ significantly in magnitude, with 2.6 representing a best-case, low-emission scenario and 8.5 representing a high-emissions, \say{worst-case} scenario} \citet[Table A.7]{kahn2021long} estimate the percent loss in GDP per capita by 2030, 2050, and 2100 for several economies: for Mexico the results indicate that under RCP 2.6 the GDP per capita increases between 0.10 and 0.23\% and under the RCP 8.5 scenario the loss is between 0.64 and 5.54\%.

The paper proceeds as follows. \Cref{sec:litrev} surveys the literature on the effect of extreme-weather related events on inflation. \Cref{sec:data}  describes the data used in the empirical analysis. \Cref{sec:method}  presents the two econometric approaches used in the empirical analysis. \Cref{sec:results} shows the results, while \Cref{sec:conclusion} offers concluding remarks.

\section{Literature review}
\label{sec:litrev}

This paper contributes to the literature on the macroeconomic effects of extreme weather-related events, with a focus on their short- and long-term impacts on GDP growth and consumer price inflation. A recent synthesis by \citet[][Table 1]{ehlers2025} summarizes the stylized and theoretical effects of such events: extreme temperatures tend to reduce economic activity without significantly affecting inflation, while precipitation and flood events are associated with short-term declines in output and increases in inflation. In the medium term, reconstruction-related capital investment may boost economic activity and offset initial inflationary pressures. The following subsections expand on these findings, highlighting studies that support or challenge this characterization.

\subsection{Short- and long-term effects of weather-related events on macroeconomic variables}

A growing body of empirical research examines the macroeconomic consequences of climate shocks, including their effects on inflation, output, fiscal balances, and long-run growth. While many studies employ the local projections framework of \cite{jorda2005estimation}, results vary significantly due to differences in shock measurement, identification strategies, data sources, sample composition, and modeling choices. A key distinction lies in the definition of climate shocks: some studies focus on discrete extreme events (e.g., disasters recorded in EM-DAT), while others analyze deviations from historical climate norms. Moreover, most contributions rely on large cross-country or global panels, estimating average effects across heterogeneous economies rather than country-specific or regional responses. These methodological choices influence the estimated magnitude, persistence, and transmission channels of climate shocks. In this paper, we adopt a deviations-based approach and focus on country-level dynamics, in contrast to event-based measures.

Many influential studies estimate average effects using global samples that pool advanced and developing economies. For example, \cite{faccia2021feeling}, \cite{kotz2023impact}, and \cite{cevik2023eye} analyze inflation responses in large cross-country panels, identifying heterogeneous effects across income groups or climate zones but reporting pooled coefficients. Similarly, disaster-based studies such as \cite{ehlers2025} and \cite{kabundi2022persistent} use internationally harmonized datasets to estimate average macroeconomic responses to extreme weather events. Long-run growth analyses by \cite{kahn2021long} and fiscal assessments by \cite{akyapi2022estimating} also rely on broad international panels. At the global time-series level, \cite{bilal2024macroeconomic} identify innovations in world temperature and map them into aggregate productivity within a neoclassical growth framework, while \cite{winter2023long} emphasize volatility and tail risks using aggregated data.

In contrast, subnational or country-level studies, such as \cite{costa2025macroeconomic} (in OCDE countries) at the regional level and \cite{liu2025macroeconomic} at the provincial level (in Canada), often uncover stronger persistence or asymmetric effects that are diluted in global panels. These analyses exploit within-country variation, holding institutional and policy frameworks constant, and can better isolate structural transmission channels. Their findings suggest that aggregation across countries may understate tail risks, sectoral vulnerabilities, or long-run growth effects associated with sustained climate deviations.

Climate shocks are measured differently across the literature. One strand focuses on temperature extremes and inflation. \cite{faccia2021feeling} show that inflationary effects depend on seasonality and shock direction, with hot summers increasing food prices—particularly in emerging markets—while medium-term responses may reverse. Using high-frequency ERA5 data in a broad cross-country panel, \cite{kotz2023impact} document nonlinear and heterogeneous effects of temperature and precipitation anomalies on month-on-month inflation, largely driven by food and varying across climate zones and income levels. Similarly, \cite{cevik2023eye} identify asymmetric effects of extreme temperatures across advanced and developing economies. While these studies emphasize heterogeneity, their estimates are typically derived from pooled global samples, implying that reported coefficients represent average effects rather than fully country-specific responses.

Another strand employs disaster-based indicators from international databases. Using EM-DAT data in a global panel, \cite{ehlers2025} show that macroeconomic effects differ sharply across disaster types: droughts generate persistent GDP losses, while many other disasters have limited aggregate effects and only short-lived inflation responses. Likewise, \cite{kabundi2022persistent} find that extreme weather shocks raise food inflation with limited aggregate persistence. At a finer spatial scale, \cite{costa2025macroeconomic} exploit regional variation within countries to estimate durable output losses from droughts using staggered-treatment local projections, whereas \cite{heinen2019price} report temporary inflationary pressures following destructive events. Compared to deviation-based approaches, disaster measures capture tail risks but may miss broader, gradual climate fluctuations; compared to country-level studies, global panels may smooth important local nonlinearities.

Long-run growth analyses also reveal substantial heterogeneity. Using an ARDL framework in a large cross-country sample, \cite{kahn2021long} find that persistent temperature deviations reduce per capita GDP growth, especially in hotter and poorer countries. In contrast, \cite{liu2025macroeconomic} focus on Canadian provinces and document asymmetric long-run growth effects of sustained temperature and precipitation deviations at the subnational level. The difference between global panels and within-country analyses highlights how aggregation can mask structural characteristics—such as sectoral composition or adaptive capacity—that shape climate sensitivity.

Fiscal consequences further illustrate the limitations of cross-country averaging. \cite{akyapi2022estimating} provide global evidence that temperature and precipitation shocks reduce GDP per capita and worsen fiscal balances, with stronger effects in economies with limited fiscal space. Yet, as in much of the literature, the estimated coefficients reflect pooled international dynamics rather than country-specific fiscal transmission mechanisms.

Identification strategies add another dimension of heterogeneity. While local projections dominate, \cite{ciccarelli2024demand} estimate a panel SVAR to distinguish physical and transition risks, and \cite{martinez2025macroeconomic} combine panel VAR methods with forecast revisions to capture expectation effects around extreme events. At the global level, \cite{bilal2024macroeconomic} identify innovations in global temperature and embed time-series estimates in a neoclassical growth framework, whereas \cite{winter2023long} emphasize volatility and tail risks. Differences in frequency, identification assumptions, and structural embedding further contribute to dispersion in estimated effects.

Overall, heterogeneity in results reflects differences in shock definitions (extremes versus deviations), climate datasets (reanalysis versus disaster databases), econometric identification, structural modeling, and the predominance of global panel estimations over country-specific or regional analyses. Inflation effects tend to be short-lived and food-driven, while output and growth responses—particularly to sustained deviations or droughts—are more persistent. These contrasts underscore the importance of distinguishing between extreme-event and deviation-based measures and between global average effects and country-level dynamics when assessing the macroeconomic consequences of climate shocks. This paper follows the analysis of \cite{faccia2021feeling}, \cite{kahn2021long}, and \cite{liu2025macroeconomic} to investigate the impact of climate deviations from historical norms on real GDP per capita and consumer price inflation in Mexico.

\subsection{Short- and long-term effects of weather-related events on Mexican macroeconomic variables}

Most studies on the effects of climate events on the Mexican economy have focused on specific sectors or markets. For instance, \cite{mendelsohn2010ricardian} use a Ricardian model and farm-level data to show that higher temperatures reduce land values, particularly for rainfed soils. Similarly, \cite{basurto2023impactos} estimate that rising temperatures could reduce income per capita in the livestock sector by up to 33 percentage points. In a region-specific study, \cite{gay2006potential} evaluate the impacts of climate change on coffee production in Veracruz and find that higher temperatures may reduce output by 34 percent.

In a more aggregated framework, \cite{arellano2023temperature} examine the effect of temperature shocks on fruit and vegetable price inflation, finding a convex U-shaped relationship: both very low and very high temperatures are associated with higher prices. Similarly, \cite{arellano2023weather_} analyze the effects of temperature and precipitation shocks on dry beans and white corn, concluding that rainfed production is particularly vulnerable to positive temperature and negative precipitation shocks—results consistent with \cite{mendelsohn2010ricardian} and \cite{basurto2023impactos}. Complementary findings are provided by \cite{estrada2022impacts} and \cite{estrada2023model}, who suggest that climate change may significantly reduce harvested areas for maize, rice, and sorghum.


This paper contributes to the literature by analyzing the effects of temperature and precipitation deviations from historical norms on real GDP per capita and consumer price inflation in Mexico. Unlike previous studies that focus on agricultural inflation, this analysis extends to headline (all-items) inflation and its components—food, services, and energy. On the production side, the study assesses impacts across the primary, secondary, and tertiary sectors. To the best of the author’s knowledge, this is the first study to apply local projections and autoregressive distributed lag (ARDL) models to estimate the macroeconomic effects of weather events in Mexico.

\section{Data}
\label{sec:data}

We built a regional panel data set that includes economic and weather-related variables, with quarterly frequency from the first quarter of 2004 to the last quarter of 2024. The empirical analysis covers seven geographic regions in Mexico: 1)~North-Centre, 2)~North-East, 3)~North-West, 4)~Northern Border, 5)~Mexico City, 6)~South, and 7)~South-Centre. The macroeconomic variables of interest in this study are headline inflation, measured as the quarterly variation of the seasonally adjusted Consumer Price Index (CPI); and total production per capita, constructed using GDP data and the 2020 population census. The climate variables, by contrast, are temperature and precipitation anomalies, measured as population-weighted monthly deviations of temperature and precipitation from their historical norms.\footnote{\,Following 
\cite{tol2017population} and \cite{kahn2021long}, we assume no population 
growth and abstract from changes in the population distribution and population 
trends. Hence, to construct population-weighted climate variables and GDP per 
capita we consider only the population level in 2020.}

\begin{figure}[!htb]
    \centering
    \subfloat[(a) Headline CPI inflation]{\includegraphics[scale=0.55]{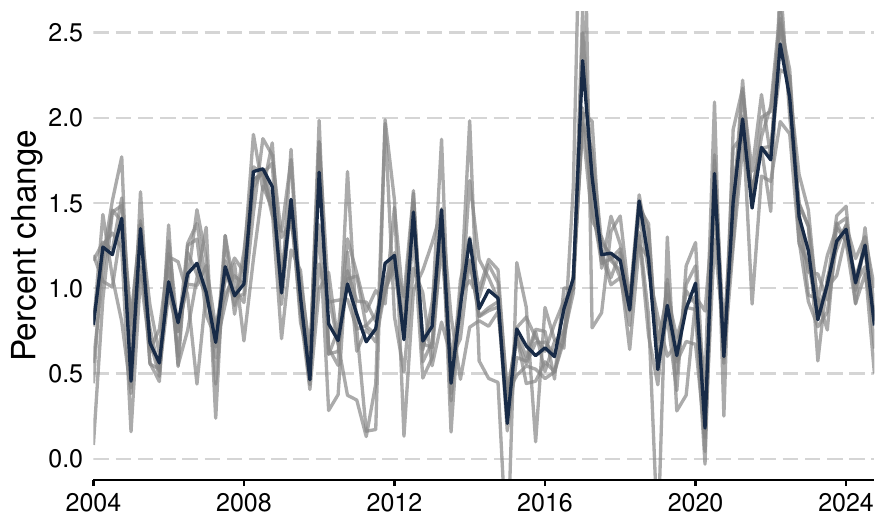}}
    \hfil
     \subfloat[(b) Total real GDP per capita]{\includegraphics[scale=0.55]{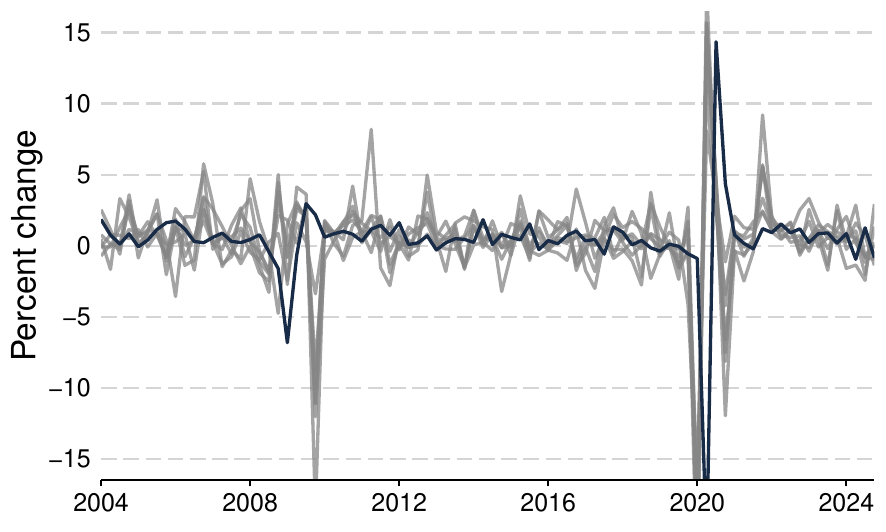}}
    \hfil
   \begin{center}
         \caption{Quarterly headline CPI inflation rate and quarterly change of total real GDP per capita}\label{fig:Region_inflation_gdp}
        \vspace{-0.15cm}
        \caption*{\small{\textbf{Source:} INEGI.\\ \textbf{Note:} The data is seasonally adjusted. Real GDP per capita is measured using 2018 Mexican pesos and the 2020 population level. The blue lines represent the national headline inflation rate and the national real GDP per capita quarterly change; the gray lines depict inflation and real GDP per capita change of the seven Mexican regions. See \autoref{fig:Region_sector_inf} and \autoref{fig:Region_sector_gdp} for a colored version of each region and for the components of inflation and real GDP per capita.}}
    \end{center}
\end{figure}

\begin{figure}[htb!]
    \centering
    \subfloat[Temperature anomaly (\unit{\celsius})]{\includegraphics[scale=0.55]{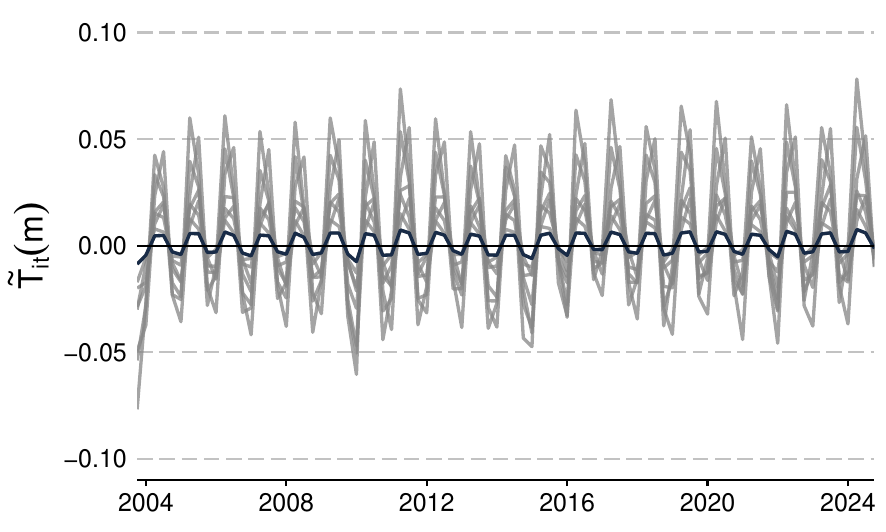}}
    \hfill
    \subfloat[Precipitation anomaly (mm)]{\includegraphics[scale=0.55]{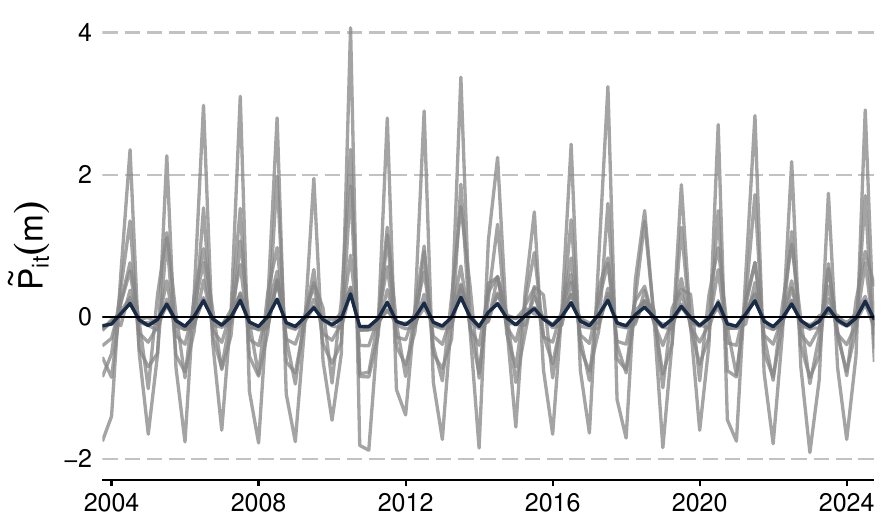}}
    \begin{center}
         \caption{Historical average of population-weighted temperature and precipitation anomalies.}\label{fig:TS_climate}
        \vspace{-0.15cm}
        \caption*{\small{\textbf{Source:} World Bank.\\ \textbf{Notes:} Temperature and precipitation anomalies are computed as the deviation from their (30-year) norm. See \autoref{eq:temp_an}.}}
    \end{center}
\end{figure}

Macroeconomic variables were retrieved from INEGI, the Mexican statistical 
agency. \autoref{fig:Region_inflation_gdp} shows the quarterly inflation rate 
and total GDP per capita in each of the seven regions in Mexico from the first 
quarter of 2004 to the fourth quarter of 2024, while \autoref{fig:TS_climate} 
shows temperature and precipitation anomalies for the same period. Although 
headline inflation and total GDP per capita are among the most closely monitored indicators of economic performance across economies, we also examine the effects of weather shocks on selected components of the CPI and GDP series. Specifically, for the CPI we consider the effects on the inflation rate of 1)~food, beverages, and tobacco; 2)~non-food goods; 3)~services; 
4)~agricultural products; and 5)~energy; while for output, we examine the GDP 
per capita of the 1)~primary, 2)~secondary, and 3)~tertiary sectors. The 
evolution of each of these components across the seven Mexican geographic 
regions is shown in \autoref{fig:Region_sector_inf} and 
\autoref{fig:Region_sector_gdp}.

Both CPI and GDP data are published by state and not by region, so we must construct the series for each region. The easiest regional series to construct are those of inflation. Inflation in region $i$, during time $t$, is represented as $\pi_{it}$, and is computed as the mean of the inflation series of the states, $s = {1,\ldots,S}$, that belong to region~$i$, that is:
\begin{equation*}\label{eq: inf_reg}
    \pi_{it} = \frac{1}{S}\sum_{s=1}^S\pi_{i,s,t},
\end{equation*}
where $\pi_{i,s,t} = \Delta\text{ln}(\text{CPI}_{i,s,t}) = \text{ln}\left(\text{CPI}_{i,s,t}\right) - \text{ln}\left(\text{CPI}_{i,s,t-1}\right)$ is the quarterly change of the seasonally-adjusted Consumer Price Index in state $s$ of region $i$, from time $t-1$ to $t$.

Regarding GDP per capita, the aggregation process from a state level into a regional level is similar to the inflation case, but with certain particularities to consider. The first consideration is that GDP by state is only publicly-published annually, which is insufficient for rigorous econometric analysis. Therefore, we address this by approximating a quarterly GDP series by state using the quarterly growth of the seasonally-adjusted Economic Activity Index of the States (ITAEE, in Spanish), which is a quarterly-published index that tracks the production growth of the Mexican states. Thus, we assume that the annual GDP reported for each state belongs to the last quarter of the year of reference and approximate the next quarter production using the first difference of ITAEE:\footnote{\,We consider the ITAEE series reported using constant 2018 prices, in order to obtain a measure of real GDP.}
\begin{equation*}\label{eq: qGDP}
    Y_{i,s,t+1} = (1 + \Delta \text{ITAEE}_{i,s,t+1}) \times Y_{i,s,t},
\end{equation*}
where $Y_{i,s,t}$ refers to the quarterly GDP of state~$s$ in region~$i$ during $t$; $\Delta$ is the rate of change~$\left(\Delta x_t= x_t/x_{t-1} - 1\right)$; and $\text{ITAEE}_{i,s,t+1}$ is the economic activity index in $t+1$. Once we have an approximation of quarterly GDP by state we must construct the GDP per capita series,~$y_{i,s,t+1}$, which is only~$Y_{i,s,t+1}$ divided by the population level in each state during 2020. Finally, the aggregation into a regional GDP per capita level is constructed using the sum of each~$y_{i,s,t}$ in~$i$:
\begin{equation*}\label{eq: qGDPpc_reg}
    y_{i,t} = \sum_{s=1}^Sy_{i,s,t} = \sum_{s=1}^S\frac{Y_{i,s,t}}{N_{i,s,2020}},
\end{equation*}
where $y_{i,t}$ is the quarterly GDP per capita level in region $i$, and $N_{i,s,2020}$ is the population-level of state $s$ in region $i$ during 2020.

Temperature and precipitation levels were obtained from the World Bank’s Climate Knowledge Portal and are measured using the ERA5 model, a state-of-the-art atmospheric reanalysis produced by the European Center for Medium-Range Weather Forecasts (ECMWF). ERA5 combines historical observations from satellites, weather stations, radio signals, and other sources with a consistent numerical weather prediction model through data assimilation techniques. The result is a global, high-resolution, gridded dataset that provides physically consistent reconstructions of past weather and climate conditions over time. Data for Mexico are available for the period 1901–2024 and for its 32 states. To construct the panel with climate anomalies, we use the time series of average temperature and average precipitation levels\footnote{\,The Portal also has available minimum and maximum temperatures and precipitation levels, however, we decided to only use the averages for the definition of climate anomaly.} and build anomalies of both weather-related variables as their deviations from their long-term average, i.e. norm, computed using a thirty-year moving average, following \cite{kahn2021long} and \cite{arguez2012noaa}.\footnote{\,We use the official World Meteorological Organization definition of climate norm and consider a thirty-year moving average for its computation. However, as in \cite{kahn2021long}, we also consider norms computed with twenty and forty-year moving averages to check the robustness of our results.}

As with some macroeconomic variables, weather-related data are reported at monthly frequency and by state. Consequently, it is transformed from monthly frequency to quarterly frequency, and later aggregated to the regional level. Provided that macroeconomic variables are generated by society's consumption decisions, we are only interested to model the effects of weather anomalies suffered by society on production and inflation, therefore, we weight the average temperature and precipitation levels obtained from the World Bank using the population level of each state and following \cite{tol2017population}:
\begin{equation*}\label{eq: weather_weighted}
    X_{i,t} = \frac{1}{\sum_{s=1}^SN_{i,s,t}}\sum_{s=1}^SX_{i,s,t}N_{i,s,t},
\end{equation*}
where $X_{i,t} $ is the temperature $T_{i,t}$ or the precipitation $P_{i,t}$ levels in region $i$ during time $t$. The term $X_{i,s,t}$ refers either to temperature or precipitation levels in state $s$ in region $i$; and $N_{i,s,2020}$ refers to the population level of state $s$ in region $i$ during 2020.

\begin{figure}
    \centering
    \subfloat[$\tilde{T}_{it}(30)$]{\includegraphics[width=0.5\textwidth]{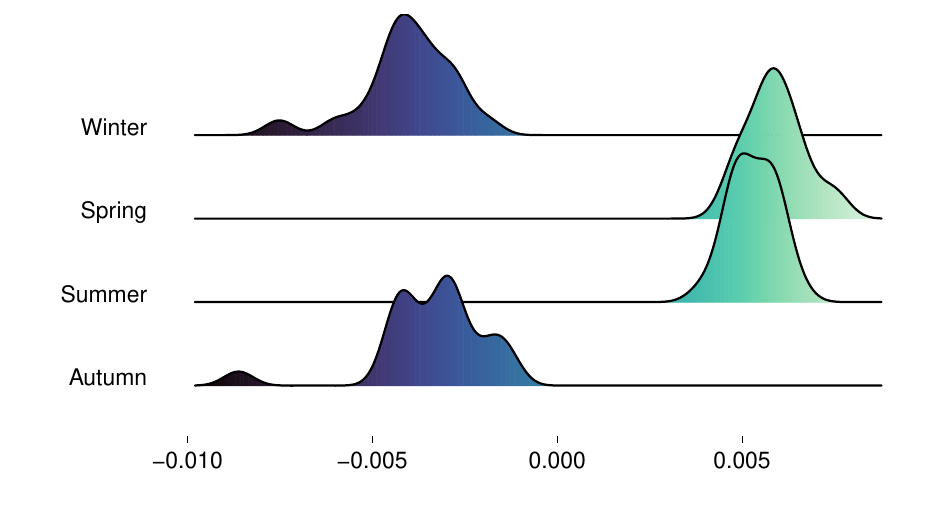}}
    \hfil
    \subfloat[$\tilde{P}_{it}(30)$]{\includegraphics[width=0.5\textwidth]{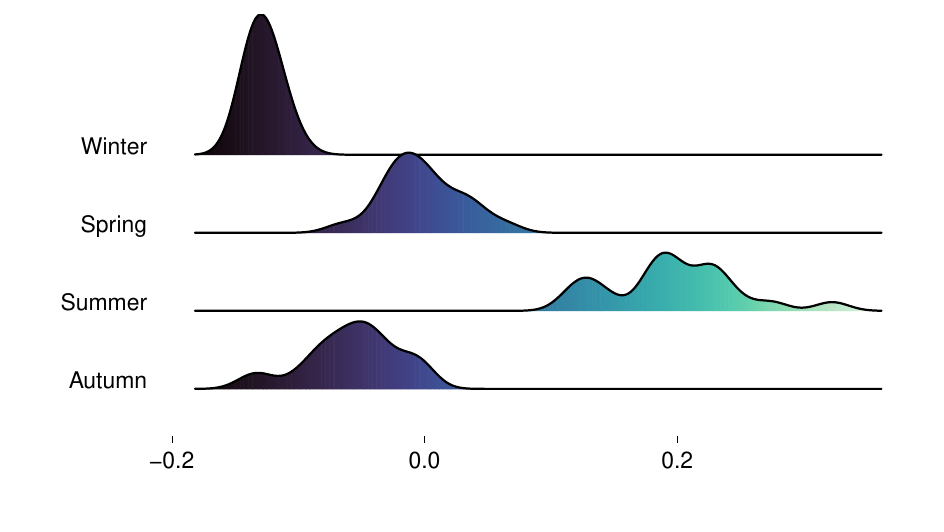}}
    
     \begin{center}
        \caption{Population-weighted weather anomalies distribution by season.} \label{fig:Nation_dist}
        \vspace{-0.15cm}
        \caption*{\footnotesize{\textbf{Source:} Own elaboration with data from INEGI.\\ \textbf{Notes:} Weather anomalies computed as deviations from their (30-year) norm. See \autoref{eq:temp_an}.}}
    \end{center}
\end{figure}

Instead of working with temperature and precipitation levels, the econometric framework is implemented using temperature and precipitation anomalies. The reason for this is that weather variables might be trended and its inclusion in the estimated specification could introduce a trend in the outcome variable that did not even exist.\footnote{\,See Appendix A.2 of \cite{kahn2021long} for a detailed analysis of how trends in temperature can lead to spurious trends in output growth in regressions used in the literature.} We follow the definition of weather anomaly of \cite{kahn2021long}, as a scaled difference between the population-weighted weather level and its population-weighted norm in the previous period:
\begin{equation}
\Tilde{X}_{it}(m) = \left(\frac{2}{m+1} \right) \left[X_{it} - X_{i,t-1}^{*}(m) \right]\,,
\label{eq:temp_an}
\end{equation}
where $i$ denotes the region, $t$ the temporal dimension, $m$ the number of years in the moving average to compute the climate norm, $X_{it}$ is the population-weighted weather-related variable, either $T_{it}$, the temperature, or $P_{it}$, precipitation levels in region $i$ at time $t$. Finally, $X_{i,t-1}^{*}(m)$ represents the historical moving average of temperature, $T_{i,t-1}^{*}(m)$, or precipitation, $P_{i,t-1}^{*}(m)$, given by 
\begin{equation*}
X_{i,t-1}^{*}(m) = \frac{1}{m}\sum_{l=1}^{m}X_{i,t-l}\,.
\end{equation*}
The terms $T_{i,t-1}^{*}(m) $ and $P_{i,t-1}^{*}(m) $ are known as time varying historical norms of temperature and precipitation, respectively. We set $m = 30$ to compute these quantities as proposed by \cite{arguez2012noaa} and used in \cite{vose2014improved} and  \cite{kahn2021long}. \autoref{fig:TS_climate} shows the population-weighted temperature (panel~a) and precipitation (panel~b)  for the seven Mexican regions. Panel~(a) shows a small positive trend for population-weighted temperature in some regions and seasonalities. Panel~(b) shows that the population-weighted precipitation time series exhibits seasonal patterns and no trends in all seven regions. \autoref{fig:Nation_dist}, \autoref{fig:Regions_temp}, and \autoref{fig:Regions_precip} show the population-weighted temperature and precipitation anomalies for Mexico and its seven regions: they show that the spring is the hottest season in Mexico and that autumn and winter are very similar, nonetheless, the winter is the season with the most extreme observations. The national summer temperature is bimodal because the hottest region differs among the regions.

\autoref{tab:descriptive_stats} presents descriptive statistics for total real GDP per capita, headline inflation, and population-weighted temperature and precipitation anomalies ($\tilde{T}_{it}(m)$ and $\tilde{P}_{it}(m)$) across eight regions of Mexico. The statistics reveal substantial regional heterogeneity in economic performance and exposure to climate variability.

In terms of economic output, Mexico City exhibits by far the highest mean real GDP per capita (366.44), more than double the national average (174.52), reflecting its role as the country's primary economic hub. The Center South has the lowest average output (114.66), indicating relatively lower regional development. In contrast, the Northern Border, North East, and North West show above-average income levels, consistent with their industrial and export-oriented economies. Notably, the dispersion of GDP per capita is highest in Mexico City (standard deviation: 30.33) and the Northern Border (22.63), while the South displays the lowest variability (6.82), suggesting greater economic stability over time.

Regarding inflation, the national average is 1.09\% with moderate dispersion (standard deviation: 0.45). Regional inflation rates are broadly similar in level, ranging from 1.02\% (Northern Border) to 1.13\% (Center South). However, the Northern Border exhibits the highest volatility (standard deviation: 0.67), indicating greater instability in price dynamics, while the South, despite having a relatively high mean inflation (1.12\%), shows lower dispersion (0.46), suggesting more stable price trends.

For temperature anomalies ($\tilde{T}_{it}(m)$), the Center South displays the highest variability (standard deviation: 0.0377), followed by the Northern Border (0.0354), indicating that these regions experience the most extreme and fluctuating temperature deviations. The Center North and North East also show relatively high dispersion, while Mexico City and the national aggregate exhibit the lowest volatility. The median values (50th percentile) are generally positive across regions, suggesting a trend toward warmer-than-average conditions over the sample period.

Precipitation anomalies ($\tilde{P}_{it}(m)$) exhibit even greater regional disparity. The Center South again stands out with the largest standard deviation (1.6498), more than nine times the national average (0.1258), reflecting highly volatile rainfall patterns. The Center North (0.8497) and Mexico City (0.4004) also experience significant variability, while the North West shows negative median precipitation anomalies (-0.1189), indicating a tendency toward drier conditions. The Northern Border and North East display moderate volatility, with distributions skewed toward negative anomalies, suggesting recurring dry spells.

Overall, the descriptive statistics highlight pronounced regional disparities in both economic conditions and climate exposure. The Center South and Northern Border emerge as the most climatically volatile regions, facing the largest fluctuations in both temperature and precipitation. These patterns underscore the importance of region-specific analyses when assessing the macroeconomic impacts of climate change in Mexico.
\begin{table}[H]
    \centering
    \caption{\small Descriptive statistics of macroeconomic and climate variables}
    \label{tab:descriptive_stats}
    \vspace{0.15cm} 
    \footnotesize 
    \begin{tabular}{llccccc}
        \toprule
         & Region & Mean & Standard deviation & Percentile 25th & Percentile 50th & Percentile 75th \\ 
  \midrule
\multirow{7}{8em}{Total real GDP per capita} & National & 174.52 & 16.85 & 160.02 & 175.54 & 188.32 \\ 
   & Center North & 166.23 & 23.44 & 144.20 & 167.07 & 188.18 \\ 
   & Center South & 114.66 & 10.14 & 106.05 & 115.55 & 123.64 \\ 
   & Mexico City & 366.44 & 30.33 & 345.17 & 366.83 & 392.22 \\ 
   & North East & 232.16 & 27.60 & 210.54 & 234.54 & 256.62 \\ 
   & North West & 181.48 & 20.22 & 165.11 & 184.46 & 201.56 \\ 
   & Northern Border & 216.17 & 22.63 & 198.88 & 214.54 & 236.85 \\ 
   & South & 149.61 & 6.82 & 147.22 & 150.79 & 153.62 \\ 
   \midrule
\multirow{7}{8em}{Headline inflation} & National & 1.09 & 0.45 & 0.78 & 1.03 & 1.36 \\ 
   & Center North & 1.12 & 0.47 & 0.79 & 1.02 & 1.38 \\ 
   & Center South & 1.13 & 0.48 & 0.75 & 1.12 & 1.42 \\ 
   & Mexico City & 1.09 & 0.42 & 0.79 & 1.04 & 1.29 \\ 
   & North East & 1.03 & 0.52 & 0.67 & 0.95 & 1.32 \\ 
   & North West & 1.04 & 0.51 & 0.70 & 0.99 & 1.39 \\ 
   & Northern Border & 1.02 & 0.67 & 0.63 & 0.94 & 1.33 \\ 
   & South & 1.12 & 0.46 & 0.75 & 1.06 & 1.44 \\ 
   \midrule
\multirow{7}{8em}{$\Tilde{T}_{it}(m)$} & National & 0.0010 & 0.0047 & -0.0035 & 0.0013 & 0.0056 \\ 
   & Center North & 0.0067 & 0.0291 & -0.0217 & 0.0047 & 0.0336 \\ 
   & Center South & 0.0104 & 0.0377 & -0.0238 & 0.0077 & 0.0423 \\ 
   & Mexico City & 0.0024 & 0.0078 & -0.0047 & 0.0025 & 0.0090 \\ 
   & North East & 0.0028 & 0.0190 & -0.0147 & 0.0043 & 0.0216 \\ 
   & North West & 0.0021 & 0.0146 & -0.0101 & 0.0027 & 0.0141 \\ 
   & Northern Border & 0.0044 & 0.0354 & -0.0292 & 0.0047 & 0.0362 \\ 
   & South & 0.0036 & 0.0133 & -0.0090 & 0.0042 & 0.0158 \\ 
   \midrule
\multirow{7}{8em}{$\Tilde{P}_{it}(m)$} & National & 0.0035 & 0.1258 & -0.1045 & -0.0267 & 0.0766 \\ 
   & Center North & 0.0150 & 0.8497 & -0.6560 & -0.2743 & 0.4660 \\ 
   & Center South & 0.0460 & 1.6498 & -1.4136 & -0.3788 & 1.1718 \\ 
   & Mexico City & 0.0114 & 0.4004 & -0.3345 & -0.1391 & 0.3615 \\ 
   & North East & 0.0081 & 0.1611 & -0.1123 & -0.0494 & 0.0888 \\ 
   & North West & -0.0047 & 0.2256 & -0.1568 & -0.1189 & 0.0299 \\ 
   & Northern Border & 0.0066 & 0.1615 & -0.1050 & -0.0475 & 0.0576 \\ 
   \bottomrule

    \end{tabular}
    \begin{minipage}{16.5cm}
        \footnotesize \textbf{Notes:} The data span from 2004Q1 to 2024Q4, comprising 84 quarterly observations.. Climate anomalies, $\Tilde{T}_{it}(m)$ and $\Tilde{P}_{it}(m)$, are computed as deviations from the (30-year) norm of temperature and precipitation, respectively (see \autoref{eq:temp_an}). Real GDP per capita is shown in thousands of Mexican pesos, and the headline inflation rate is the quarterly rate of change of the Consumer Price Index expressed as a percentage. $\Tilde{T}_{it}(m)$ and $\Tilde{P}_{it}(m)$ are in  degrees Celsius (°C) and millimeters (mm), respectively.
    \end{minipage}
\end{table}

\section{Methodology}\label{sec:method}

There are two leading empirical methodologies used to study the effects of weather anomalies on GDP per capita and CPI growth. To assess the effects on GDP, we estimate the impact using a function of the parameters from the autoregressive distributed lag (ARDL) model as in \cite{kahn2021long} and \cite{liu2025macroeconomic}. To analyze the impact on inflation, we estimate the impulse-response function (IRF) using the panel local projection model of \cite{jorda2005estimation}, closely following the specification proposed by \cite{faccia2021feeling} and \cite{ehlers2025}.\footnote{\, Rather than measuring temperature and precipitation anomalies as dichotomous variables indicating whether deviations from their long-term norms exceed a given threshold during specific seasons, we define climate-related anomalies as in \autoref{eq:temp_an}. Moreover, we use a moving-average climate norm—based on 20-, 30-, and 40-year windows—to construct temperature and precipitation anomalies, rather than a fixed norm as in \cite{faccia2021feeling}. This dynamic approach accounts for potential trends in the temperature and precipitation time series \citep{kahn2021long}.}

\subsection{Measuring effects of climate anomalies on economic growth}\label{sec:method_ardl}

We employ a panel autoregressive distributed lag (ARDL) specification similar to \cite{kahn2021long} to estimate the long-run effects of weather shocks on the GDP per capita. The methodology used for this purpose delivers an estimation of the long-run effects of climate anomalies, by approximating the steady-state equilibrium of the theoretical macroeconomic model developed by \cite{binder1999stochastic}, and extended by \cite{kahn2021long} to account for climate-related productivity shocks. The macroeconomic model is a stochastic extension of the Solow–Swan growth framework in which technological progress follows an exogenous stochastic process.\footnote{\, See online appendix A of \cite{kahn2021long} for the full derivation of the model.} 

Therefore, to estimate the long-run effects of weather anomalies, we first approximate the steady-state value of production per capita,~$y_{i,t}$, by a linear stationary process with possibly common factors following the approach of \cite{kahn2021long} and considering the ARDL framework discussed in \cite{Pesaran_Shin_1999} and \cite{pesaran2001bounds}. We estimate, thus, the following panel specification for $y_{i,t}$:
\begin{eqnarray}\label{model:ardl}
    \Delta \text{ln}y_{i,t} = \alpha_{i} + \sum_{l=1}^{p} \varphi_{l}\Delta \text{ln}y_{i,t-l} + \sum_{l=0}^{q}\bm{\beta}_{l}^{\top} \Delta \tilde{\vx}_{i,t-l}(m) + \varepsilon_{i,t}\,,
\end{eqnarray}
where $\Delta$ denotes the difference operator~($\Delta x_{t} = x_{t} - x_{t-1}$); $y_{i,t}$ represents the level of production (GDP) per capita in region $i$ at time $t$; $\alpha_{i}$ refers to region fixed effects; ${\tilde{\vx}_{i,t}(m) = \left[ \Tilde{T}_{i,t}^{+}(m),\Tilde{T}_{i,t}^{-}(m),\Tilde{P}_{i,t}^{+}(m),\Tilde{P}_{i,t}^{-}(m)\right]}$ is a matrix containing our measures of weather anomalies: temperature and precipitation deviations from their norms, computed with $m=30$;\footnote{\,In order to check for the robustness of our results, we also estimate the model considering climate anomalies computed with norms of $m=20$ and $m=40$ years.} $\varepsilon_{i,t}$ denotes the error term, while $\varphi$ and $\beta$ are parameters to estimate. Note that, in contrast to \autoref{sec:method_lp}, both positive and negative deviations from the norm are considered in $\tilde{\vx}_{i,t}(m)$, as this allows us to account for the possible asymmetry regarding the effects that, for instance, colder or hotter temperatures might have on growth \citep{kahn2021long}. 

The average long-term effect of weather anomalies on production per capita,~$\theta$, is computed from the estimated short-term parameters of \autoref{model:ardl},~$\hat{\bm{\beta}}$ and~$\hat{\varphi}$. Accordingly, the estimated long-run effect of climate anomalies reflects the cumulative marginal impact of past and current weather anomalies on production per capita, scaled by the speed at which Mexican regions converge to their steady state. Formally, the long-term effect is given by~$\hat{\theta} = \hat{\phi}^{-1} \sum_{l=0}^{p}\hat{\bm{\beta}}_{l}$, where~$\hat{\phi} = 1 - \sum_{l=1}^{p}\hat{\varphi}_{l}$ is the error-correction term,\footnote{\,The asymptotic standard errors associated to the long-run estimates of the model are computed using the Delta method, as suggested by \cite{pesaran2015time}.} interpreted in this framework as the region’s adjustment speed towards its steady state.

As in \cite{kahn2021long}, we contend that the ARDL framework proposed by \cite{Pesaran_Shin_1999} and \cite{pesaran2001bounds}  is particularly well suited for estimating long-run dynamics between macroeconomic variables and climate indicators. 
The ARDL approach accommodates serial correlation through flexible lag-length selection, yields consistent estimates of long-run parameters, and can be efficiently estimated using ordinary least squares (OLS). Moreover, as noted by \cite{cho2023recent}, the ARDL specification admits a straightforward error-correction representation with a clear economic interpretation, which in this study captures the speed at which Mexican regions converge to their steady state.\footnote{\,For a thorough discussion of the advantages, advances, and modifications to ARDL models, see \cite{cho2023recent} and the references therein. For an introduction to the model, see \cite{pesaran2015time}.}

Regarding the details of the ARDL specification in \autoref{model:ardl}, note that prices are not considered in the empirical long-run analysis. In the model of \cite{binder1999stochastic}, the steady state is defined as a balanced growth path in which per capita real variables converge to constant values (or grow at a constant exogenous rate), conditional on the stochastic process governing technology. Within this theoretical structure, steady-state outcomes are fully characterized by real variables: the capital accumulation equation and the production function, which jointly determine long-run output per capita, depend exclusively on real quantities and technological parameters. Prices are taken as exogenous and do not enter either the law of motion for capital or the equilibrium conditions that characterize the steady state, nor do they affect capital deepening or long-run productivity. Consequently, prices do not exhibit meaningful steady-state dynamics and cannot transmit permanent effects of weather shocks. Including them in the long-run empirical analysis would therefore be inconsistent with the underlying theory and would not yield interpretable estimates of long-term economic effects. For this reason, the long-run analysis is restricted to production per capita, the variable for which the model delivers well-defined and economically meaningful steady-state implications.

As a result, weather shocks can affect long-run outcomes only through their impact on real productivity and capital accumulation, generating permanent effects on output per capita. In contrast, prices may respond to shocks in the short run but lack a structural channel through which they can influence steady-state allocations or growth rates. Consequently, prices do not possess a well-defined long-run equilibrium within the model and cannot exhibit persistent responses consistent with steady-state shifts. Including prices in the long-run ARDL estimation would, therefore, violate the model’s identifying assumptions and risk conflating transitory nominal adjustments with genuine long-run real effects. For these reasons, the empirical long-run analysis focuses exclusively on production per capita, which is the sole variable for which the model provides a coherent and theoretically grounded notion of long-run persistence.

\subsection{Measuring effects of climate anomalies on inflation}\label{sec:method_lp}

The immediate effects of climate shocks on macroeconomic variables such as inflation  can be estimated using impulse response functions (IRFs). An IRF is defined as the difference between the expected value of a macroeconomic variable conditional on a nonzero climate shock  and its expected value conditional on the absence of such a shock, that is,
\begin{equation*}
    \text{IRF}(h,\delta) = \Exp(y_{i,t+h}|\varepsilon_t=\delta, \varepsilon_{t+1}=\ldots=\varepsilon_{t+h}=0) - \Exp(y_{i,t+h}|\varepsilon_t=\varepsilon_{t+1}=\ldots=\varepsilon_{t+h}=0),
\end{equation*}
where $h=1,\ldots,H$ refers to the horizon of the response, and $\delta$ refers to the size of the climate shock, which we will assume to be equal to one. To identify the shocks, and therefore the impulse-response functions, we use the local projection methodology of \cite{jorda2005estimation}, following the specification proposed by \cite{faccia2021feeling}, in which accumulated inflation over time is explained by past values of the first-difference of the price index logarithm and by the time difference of climate anomalies, where climate anomalies are measured as deviations from their thirty-year norm. The IRFs are estimated from the following model:
\begin{eqnarray}\label{eq: lp_model}
    \text{ln}(y_{i,t+h}) - \text{ln}(y_{i,t-1}) = \beta^{h}\Delta\Tilde{\vx}_{i,t}(m) + \sum_{l=1}^{8} \gamma_{l}^{h}\Delta \text{ln}(y_{i,t-l}) + \alpha_{i}^{h} + \theta_{t}^{h} + \varepsilon_{i,t}^{h}\,,
\end{eqnarray}
where $\Tilde{\vx}_{i,t}(m)=\left[\Tilde{T}_{i,t}(m), \Tilde{P}_{i,t}(m)\right],$~refers either to temperature or precipitation anomalies, measured as the deviations from their norms of $m=30$ years; $\Delta$~denotes the difference operator~($\Delta x_{t} = x_{t} - x_{t-1}$), while $\alpha_{i}^{h}$~and~$\theta_{t}^{h}$ denote, respectively, region and time fixed effects; and $\varepsilon_{i,t}^{h}$~is an error term. The dependent variable is the cumulative growth rate of the natural logarithm of the CPI,~$\text{ln}(y_{i})$, of each region between horizons $t+h$ and $t-1$. Regressions are estimated for each horizon $h=0, 1, \ldots, 8$ to capture the contemporaneous effect as well as the impact over the subsequent two years (eight quarters). The~$\beta^h$ and the~$\gamma_l^h$,~$l = 1,\ldots,8$, are parameters to estimate, where the~$\beta^h$ is the parameter of interest. The aim of using fixed effects is to control for important region-invariant factors, such as latitude, and time-specific phenomena unrelated to climate. The effect of climate anomalies on macroeconomic variables $h$-periods ahead, with a shock size of $\delta=1$ is estimated with the following impulse-response function,
\begin{equation*}\label{eq:IRF_lp}
    \text{IRF}(h,1)=\beta^h\,.
\end{equation*}
As in \cite{faccia2021feeling}, we treat within-region temperature fluctuations as exogenous and prefer the panel LP model \`a la \cite{jorda2005estimation} over the VAR due to its advantages; for instance, the IRFs are easily obtained from a model which is robust to misspecification of the data-generating process (DGP) and is estimated using ordinary least squares. In addition, the IRFs estimated from the LP model are exactly the same in population as those estimated from a VAR model \citep{jorda2005estimation, plagborg2021}. 

Note that in \autoref{eq: lp_model}, we include eight lags of the first difference of the log CPI. This choice is motivated by the requirement that, for the robustness properties of LPs to hold, the specification must control for those lagged values of the data that are strong predictors of the dependent variable \citep{olea2024double}. More generally, when estimating impulse responses at horizon $h$, it is necessary to condition on at least $h$ lags of the relevant variables in order to purge the error term of predictive information contained in past realizations and to ensure that the estimated response captures the causal effect of the shock rather than omitted dynamic dependence \citep{Baumeister2025}. Consequently, we control for quarterly inflation up to eight quarters in the past, matching the maximum impulse-response horizon considered in our analysis. This alignment between lag length and horizon is essential to avoid dynamic misspecification and to maintain the interpretation of the LP coefficients as impulse responses. 

Since the inclusion of multiple lags induces serial correlation in the regression residuals by construction, the error term in our specification is subject to autocorrelation. To address this issue, we employ \cite{DriscollKraay} standard errors, which are robust to heteroskedasticity, serial correlation, and cross-sectional dependence. Given that our analysis focuses on aggregate rather than region-specific inflation dynamics, the use of these robust standard errors further shields our inference from potential cross-sectional dependence across Mexican regions.

An alternative approach is to use a Vector Autorregresion (VAR) model. VAR models offer several advantages over LPs for impulse response analysis. In particular, VARs impose internally consistent dynamic restrictions across horizons, yielding smoother and more economically coherent impulse responses. By exploiting cross-equation restrictions, VARs achieve greater statistical efficiency and typically produce tighter confidence bands than LPs of comparable order. Their framework is especially well suited for long-horizon analysis when lag lengths are extended and disciplined through shrinkage techniques. Moreover, VARs align closely with structural macroeconomic models, making them a natural bridge between theory and empirics\footnote{\,The view that VARs align closely with structural macroeconomic models is well established in the literature. Since the seminal contribution of \cite{Sims1980}, VARs have been understood as system-based representations of macroeconomic dynamics that admit economically meaningful structural interpretations through identification restrictions. A large literature shows that linearized DSGE models imply finite- or infinite-order VAR representations, making VARs natural empirical approximations to structural economic systems \citep{SmetsWouters2003, Giacomini2013}. The DSGE-VAR framework developed by \cite{DelNegroSchorfheide2004, DelNegroSchorfheide2006} formalizes this link by embedding theoretical cross-equation restrictions into VARs via priors, allowing researchers to assess the empirical fit of structural models while relaxing their restrictions. As a result, structural VARs have become a standard empirical tool for confronting macroeconomic theory with the data, particularly through impulse response analysis and structural shock identification \citep{StockWatson2001, Lutkepohl2005}.}. Finally, VARs allow for a richer and more transparent treatment of identification uncertainty by incorporating economic priors directly into the estimation process \citep{Baumeister2025}. 

Given that our framework relies solely on the assumption of weak stationarity \citep{plagborg2021}, we accommodate a broad class of potential data-generating processes, acknowledging that the true DGP is unknown and that no structural model has yet achieved general acceptance as its accurate representation. Allowing for this degree of generality necessarily implies wider confidence intervals and higher estimation variance. When the model is potentially misspecified, fitting a finite-order VAR may yield biased impulse response estimates and confidence intervals that are poorly centered \cite{olea2024double}. In this environment, results such as the finite-sample equivalence between LPs and VARs established by \cite{Ludwig2024} cannot be directly invoked, as they rely on linearity of the underlying DGP, whereas a growing body of empirical evidence suggests that the weather-macroeconomic dynamics may exhibit important nonlinearities \citep{DellJonesOlken2012, BurkeHsiangMiguel2015, ZhangDeschenesMeng2017, faccia2021feeling, kahn2021long, LiCongGuZhang2021, DongTolWang2025}. Moreover, since our analysis focuses on short-term rather than long-term effects, the advantages of imposing the dynamic structure required for VAR-based extrapolation are limited. Consequently, accepting higher estimation variance in exchange for reduced bias represents a natural and well-justified tradeoff, making LPs an attractive choice for this work, where robustness to misspecification and short-horizon inference are primary concerns.

\section{Results}\label{sec:results}
In this section, we present the effects of weather anomalies on GDP per capita and in the inflation rate using the methodology described in \autoref{sec:method}.

\subsection{The effects of weather anomalies on real GDP per capita growth}
\label{sec:ARDL}
Consistent with the existing literature, the real GDP per capita growth effects were estimated using climate anomalies computed as the difference from the 30-year climate norm, however, to ensure the robustness of our results, we also performed the analysis with anomalies computed using 20- and 40-year climate norms. As specified in \autoref{model:ardl}, we distinguish between positive and negative deviations from climate norms, and perform the estimations using a panel of seven Mexican regions. Following \cite{kahn2021long}, estimating separate effects between positive and negative deviations allows us to account for potential asymmetries in the effects of climate anomalies, that is, it might be that in some regions of the country lower or higher temperatures or precipitations might in fact, through their effects on productivity, increase the level of real GDP per capita. 

The fixed effects estimates of the long-run impact of climate anomalies on real GDP per capita growth~($\hat\theta$) and the estimated coefficients of the error correction term~($\hat\phi$) are shown in \autoref{tab:rest_all_ardl}. The table reports results for three different specifications: (i)~Specification~1 includes both positive and negative deviations of temperature and precipitation; (ii)~Specification~2 includes only positive and negative temperature deviations; and (iii)~Specification~3 considers only positive and negative precipitation deviations. Hereafter, we focus on the results for \(m=30\). We use the conventional~5\% significance level as a rule of thumb to determine whether an effect is statistically significant.  \autoref{tab:sectors_ARDL} shows estimates of the shocks of the climate variables on real GDP per capita by productive sector: primary, secondary, and tertiary. 

The results in \autoref{tab:rest_all_ardl} indicate that there is no statistical evidence of long-term effects of climate anomalies. Although the signs reported in the table coincide with some estimated in the literature \citep{kahn2021long, akyapi2022estimating, bilal2024macroeconomic}, in reality we cannot statistically distinguish them from zero. Note that none of the three specifications reported in the table indicate any long-run effects on total real GDP growth after a shock on climate anomalies. The only estimations that we can rule out to be equal to zero are those of the error correction-term, which indicates the speed of convergence of the Mexican real GDP per capita growth rate to its steady-state level. 

\begin{table}[htb]
    \centering
    \caption{\small Long-run effects of positive and negative weather anomalies on real GDP per capita growth.}
    \label{tab:rest_all_ardl}
    \vspace{0.15cm} 
    \scriptsize 
    \begin{tabularx}{\textwidth}{mmmmmmmmmmm}
        \toprule
        & \multicolumn{3}{c}{Specification 1} & \multicolumn{3}{c}{Specification 2} & \multicolumn{3}{c}{Specification 3} \\
         \cmidrule(lr){2-4} \cmidrule(lr){5-7} \cmidrule(lr){8-10}
         & m = 20 & m = 30 & m = 40 & m = 20 & m = 30 & m = 40 & m = 20 & m = 30 & m = 40 \\ 
  \midrule
$\hat{\theta}_{\Delta\Tilde{T}_{it}(m)^{+}}$ & -3.197  & -4.775  & -6.103  & -2.204  & -3.312  & -4.107  &  &  &  \\ 
   & (1.998) & (3.015) & (3.954) & (1.466) & (2.226) & (2.906) &  &  &  \\ 
  $\hat{\theta}_{\Delta\Tilde{T}_{it}(m)^{-}}$ & -0.872  & -1.334  & -1.700  & -0.817  & -1.299  & -1.638  &  &  &  \\ 
   & (1.222) & (1.803) & (2.416) & (1.106) & (1.609) & (2.163) &  &  &  \\ 
  $\hat{\theta}_{\Delta\Tilde{P}_{it}(m)^{+}}$ & 0.021  & 0.032  & 0.041  &  &  &  & 0.029  & 0.045  & 0.057  \\ 
   & (0.025) & (0.036) & (0.048) &  &  &  & (0.019) & (0.028) & (0.037) \\ 
  $\hat{\theta}_{\Delta\Tilde{P}_{it}(m)^{-}}$ & 0.064  & 0.091  & 0.124  &  &  &  & 0.060  & 0.086  & 0.117  \\ 
   & (0.076) & (0.113) & (0.151) &  &  &  & (0.072) & (0.106) & (0.142) \\ 
  $\hat{\phi}$ & 1.745 *** & 1.743 *** & 1.741 *** & 1.804 *** & 1.806 *** & 1.807 *** & 1.775 *** & 1.774 *** & 1.774 *** \\ 
   & (0.323) & (0.322) & (0.321) & (0.347) & (0.351) & (0.351) & (0.337) & (0.336) & (0.336) \\ 
   \bottomrule

    \end{tabularx}
    \begin{minipage}{16.5cm}
        \footnotesize \textbf{Notes:} The estimation is made using a panel of seven Mexican regions, with population-weighted climate, and macroeconomic data from the first quarter of 2000 to the fourth quarter of 2024. The estimated long-run effects, $\hat\theta$, are calculated from the short-run OLS estimates of \autoref{model:ardl}, that is, $\hat{\theta} = \hat{\phi}^{-1} \sum_{l=0}^{p}\hat{\bm{\beta}}_{l}$, where~$\hat{\phi} = 1 - \sum_{l=0}^{p}\hat{\varphi}_{l}$. Cross-sectional and serial correlation in the data were addressed by using  \cite{DriscollKraay} standard errors in the estimations of the short-run parameters, $\hat{\beta}$ and $\hat{\varphi}$. The standard errors associated to the long-run estimations of $\theta$ and $\phi$ are computed with the Delta Method, as suggested by \cite{pesaran2015time}, and are shown in parentheses. Asterisks indicate statistical significance at 1\%~(***), 5\%~(**), and 10\%~(*) levels. Temperature (precipitation) is measured in degrees Celsius (millimeters). The letter $m$ stands for the years used for the construction of the climate norm.
    \end{minipage}
\end{table}

However, some estimations reported in \autoref{tab:sectors_ARDL} do show some significant effects. The productive sector whose real GDP per capita growth does suffer from long-term effects after shocks on weather anomalies is the primary sector. Note that positive and negative temperature anomalies have both an effect different from zero only when controlling for precipitation deviations, as in specification~1. If precipitation is not controlled for, such as in specification~2, only negative deviations produce long-term effects on real GDP per capita growth. However, in both specifications, the effect on real GDP per capita growth is negative. That is, after controlling for precipitation deviations, the long-term effect of an annual increase of 0.01°C in the temperature above its historical norm, will reduce real GDP per capita growth in the primary sector by~0.002~percentage points a year.\footnote{\,Calculated as $0.002 = \frac{2}{m + 1}\times(-3.734*0.01)$, where $m = 30$. The long-term estimate was multiplied by 0.01 because there are no big enough temperature deviations (1°C), in absolute value, in the data.} In specification~3 of \autoref{tab:sectors_ARDL}, we note that the long-term effect of an increase of one millimeter in the precipitation above its historical norm, will increase the real GDP per capita growth in the secondary sector by 0.004 percentage points a year.\footnote{\,The 0.004 percentage points increase on real secondary sector GDP per capita growth of precipitation anomalies where computed similarly to those of temperature anomalies, however, we do not multiply by 0.01 the long-term estimate.} Note that this is the only case where asymmetries matter for long-term effects.

The results reported in \autoref{tab:rest_all_ardl} are similar to those found in a global scale by \cite{kahn2021long}, in the sense that there are no, or little evidence, of asymmetries in the long-run relationship between real GDP per capita growth and climate deviations from their historical norms. However, the causes of the similarity of our results to those of \cite{kahn2021long}, for instance, have different origins. Even though both results reject the existence of asymmetries due to the similarity of the magnitude of the estimates of $\Tilde{T}_{it}(m)^{-}$ and $\Tilde{P}_{it}(m)^{-}$, our results differ due to the lack of statistical significance of the long-term effects of temperature deviations, with the exception of the categories described above. As in \cite{kahn2021long}, we therefore re-estimate \autoref{model:ardl} with a much simpler specification. We now use the effect of $|\Tilde{T}_{it}(m)^{-}|$ and $|\Tilde{P}_{it}(m)^{-}|$ and focus on the magnitude of the deviations on real GDP growth. The results are reported in \autoref{tab:rest_all_ardl_abs}.

\begin{table}[htb]
    \centering
    \caption{\small Long-run effects of absolute weather anomalies on real GDP per capita growth.}
    \label{tab:rest_all_ardl_abs}
    \vspace{0.15cm} 
    \scriptsize 
    \begin{tabularx}{\textwidth}{mmmmmmmmmmm}
        \toprule
        & \multicolumn{3}{c}{Specification 1} & \multicolumn{3}{c}{Specification 2} & \multicolumn{3}{c}{Specification 3} \\
         \cmidrule(lr){2-4} \cmidrule(lr){5-7} \cmidrule(lr){8-10}
         & m = 20 & m = 30 & m = 40 & m = 20 & m = 30 & m = 40 & m = 20 & m = 30 & m = 40 \\ 
  \midrule
$\hat{\theta}_{\Delta|\Tilde{T}_{it}(m)|}$ & -1.204  & -1.827  & -2.438  & -1.240  & -1.900  & -2.515  &  &  &  \\ 
   & (1.278) & (1.909) & (2.538) & (1.265) & (1.867) & (2.471) &  &  &  \\ 
  $\hat{\theta}_{\Delta|\Tilde{P}_{it}(m)|}$ & 0.039  & 0.056  & 0.072  &  &  &  & 0.046  & 0.067  & 0.088  \\ 
   & (0.026) & (0.038) & (0.050) &  &  &  & (0.032) & (0.048) & (0.062) \\ 
  $\hat{\phi}$ & 1.781 *** & 1.782 *** & 1.778 *** & 1.851 *** & 1.851 *** & 1.847 *** & 1.768 *** & 1.767 *** & 1.767 *** \\ 
   & (0.333) & (0.333) & (0.330) & (0.339) & (0.340) & (0.339) & (0.361) & (0.360) & (0.359) \\ 
   \bottomrule

    \end{tabularx}
    \begin{minipage}{16.5cm}
        \footnotesize \textbf{Notes:} The estimation is made using a panel of seven Mexican regions, with population-weighted climate, and macroeconomic data from the first quarter of 2000 to the fourth quarter of 2024. The estimated long-run effects, $\hat\theta$, are calculated from the short-run OLS estimates of \autoref{model:ardl}, that is, $\hat{\theta} = \hat{\phi}^{-1} \sum_{l=0}^{p}\hat{\bm{\beta}}_{l}$, where~$\hat{\phi} = 1 - \sum_{l=0}^{p}\hat{\varphi}_{l}$. Note that, in this case, $\tilde{\vx}_{i,t}(m) = \left[ |\Tilde{T}_{i,t}(m)|,|\Tilde{P}_{i,t}(m)|\right]$ in \autoref{model:ardl}. Cross-sectional and serial correlation in the data were addressed by using  \cite{DriscollKraay} standard errors in the estimations of the short-run parameters, $\beta$ and $\varphi$. The standard errors associated to the long-run estimations of $\theta$ and $\phi$ are computed with the Delta Method, as suggested by \cite{pesaran2015time}, and are shown in parentheses. Asterisks indicate statistical significance at 1\%~(***), 5\%~(**), and 10\%~(*) levels. Temperature (precipitation) is measured in degrees Celsius (millimeters). The letter $m$ stands for the years used for the construction of the climate norm.
    \end{minipage}
\end{table}

Note that the origin of the similarity of our results to those of \cite{kahn2021long} mattered for the results obtained in \autoref{tab:rest_all_ardl_abs}, especially with respect to temperature anomalies, as it shows that neither temperature nor precipitation deviations have an effect on the growth rate of real GDP per capita. Again, only the steady-state convergence coefficient is significant. In table \autoref{tab:sectors_ARDL_abs}, temperature anomalies do show long-term effects on the real GDP per capita growth rate in the primary sector. The results are significant in both specification~1 and specification~2, and their estimated long-term coefficients are all negative, indicating that an increase in the temperature above its historical norm of 0.01°C, will reduce the real GDP per capita growth rate in the primary sector. In average, the temperature anomalies shocks will decrease real GDP growth by 0.002 percentage points by year.

\subsection{The effects of weather anomalies on inflation}
\label{sec:LPs}

\begin{figure}[htb!]
    \centering
    \subfloat[(a) All items]{\includegraphics[width=0.33\textwidth]{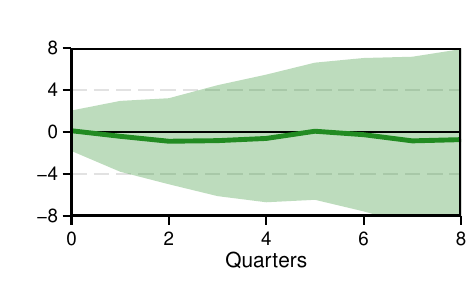}}
    \hfil
    \subfloat[(b) Services]{\includegraphics[width=0.33\textwidth]{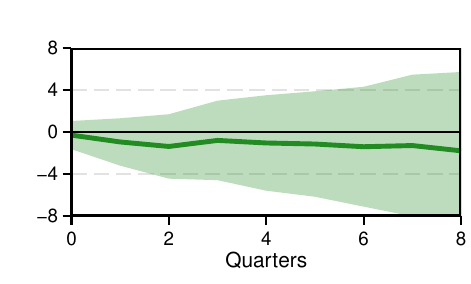}}
    \hfil
    \subfloat[(c) Non-food goods]{\includegraphics[width=0.33\textwidth]{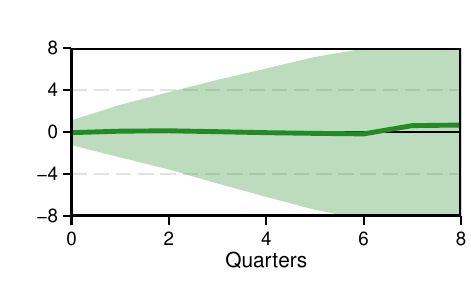}}
    \hfil
    \subfloat[(d) Food, beverages,and tobacco]{\includegraphics[width=0.33\textwidth]{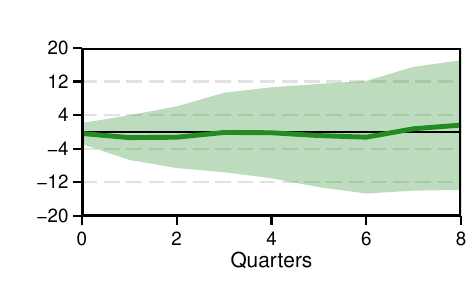}}
    \hfil
    \subfloat[(e) Agriculture]{\includegraphics[width=0.33\textwidth]{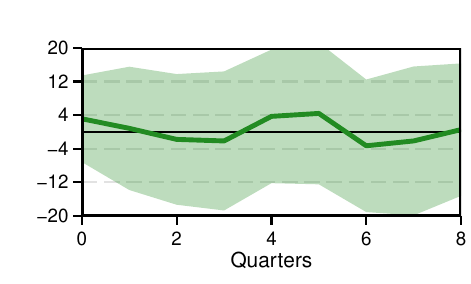}}
    \hfil
    \subfloat[(f) Energy]{\includegraphics[width=0.33\textwidth]{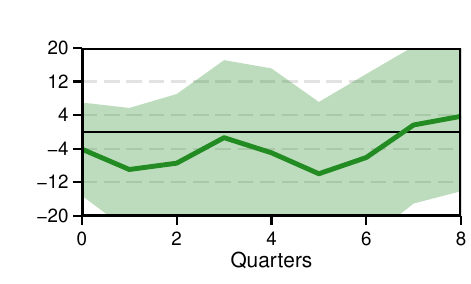}}
    \begin{center}
        \caption{Effect of temperature anomalies shocks on cumulative CPI and its components.}\label{fig:LP_INPC-tempdev}
        \vspace{-0.15cm}
        \caption*{\footnotesize{\textbf{Notes:} Cumulative impulse-responses to shocks on population-weighted temperature anomalies in seven regions of Mexico. The shaded area represents 95\% confidence intervals using \cite{DriscollKraay} standard errors.}}
    \end{center}
\end{figure}

\begin{figure}[htb!]
    \centering
    \subfloat[(a) All items]{\includegraphics[width=0.33\textwidth]{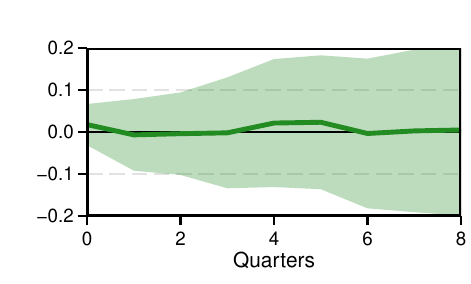}}
    \hfil
    \subfloat[(b) Services]{\includegraphics[width=0.33\textwidth]{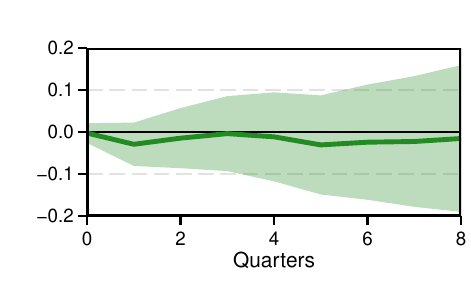}}
    \hfil
    \subfloat[(c) Non-food goods]{\includegraphics[width=0.33\textwidth]{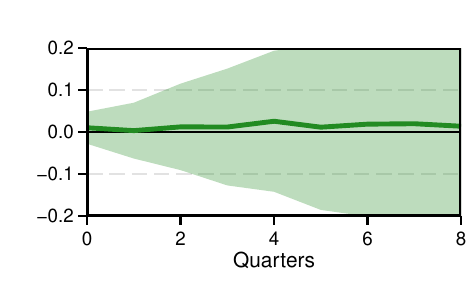}}
    \hfil
    \subfloat[(d) Food, beverages,and tobacco]{\includegraphics[width=0.33\textwidth]{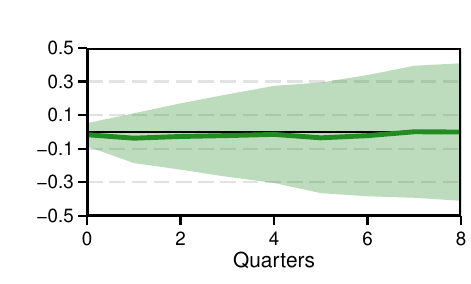}}
    \hfil
    \subfloat[(e) Agriculture]{\includegraphics[width=0.33\textwidth]{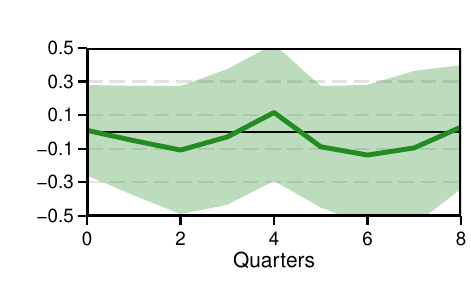}}
    \hfil
    \subfloat[(f) Energy]{\includegraphics[width=0.33\textwidth]{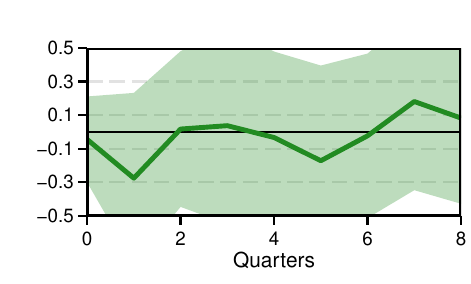}}
    \begin{center}
        \caption{Effect of precipitation anomalies shocks on cumulative CPI inflation and its components.}\label{fig:LP_INPC_precipdev}
        \vspace{-0.15cm}
        \caption*{\footnotesize{\textbf{Notes:} Cumulative impulse-responses to population-weighted precipitation deviations from its historical norm in seven regions of Mexico. The shaded area represents 95\% confidence intervals using \cite{DriscollKraay} standard errors.}}
    \end{center}
\end{figure}

In \autoref{sec:method_lp} we  noted that under our framework, long-run effects on prices are not easily identified. Therefore, to analyze the possible effects of weather anomalies on inflation, we follow \cite{faccia2021feeling} and estimate impulse-response functions à la \cite{jorda2005estimation}. Instead of estimating long-term effects, we now describe short and medium term dynamics. \autoref{fig:LP_INPC-tempdev} and \autoref{fig:LP_INPC_precipdev} report the response of the headline (all items) inflation rate, and four components of the Consumer Price Index, to a one-unit shock to temperature and precipitation deviations from their historical norm, respectively. In contrast to the previous section, our main results are not estimated controlling for any asymmetry in climate anomalies; we therefore presentt the impulse-response functions implied by \autoref{eq: lp_model}.

Note that in both figures, neither inflation measure respond to weather anomalies shocks. The agriculture inflation impulse-response indicate that climate deviations from the historical norm only has long-term effects, and that these only pass-through the real sector by its effects on productivity, as inflation remains unaltered after the shock. These results might suggest that climate shocks are only captured by changes in relative prices rather than in the price level, or that there are enough substitutes in the market such that any weather event might not cause enough scarcity to alter the price level.\footnote{\, Recall that in the present analysis, extreme-weather events such as hurricanes, earthquakes, and droughts, are not taken into account as they imply a long-tailed distribution that does not make the assumptions of our models to hold, thereby violating the assumptions required for econometric identification.} Further, it may be the case that actual asymmetry in the responses to positive and negative climate deviations drive our results to zero.

Therefore, as in the previous section, we test whether any asymmetry exists and if it does, to which extent drives the price response to climate shocks. In \autoref{tab:rest_inpc_deviationsLP}, we report the response of the same inflation series of \autoref{fig:LP_INPC-tempdev} and \autoref{fig:LP_INPC_precipdev} to a unit impulse to positive and negative temperature and precipitation deviations from their historical norms. As before, we rule out any effect of climate anomalies on inflation, as asymmetries do not exist, and therefore, cannot drive our main results. Thus, we believe that if there are any effects on prices it might only be on very specific sectors or geographical areas of the country. \autoref{tab:rest_gdp_deviationsLP} and \autoref{fig:LP_GDP-dev} show the responses of total real GDP per capita and by sectors, to the same impulse to climate anomalies as the inflation series. Again, we find no significant effect of temperature and precipitation anomalies, even after controlling for possible asymmetries.
 
\section{Conclusion}
\label{sec:conclusion}

 In this paper, we study the effects of temperature and precipitation anomalies on Mexican GDP per capita and inflation. We employ two leading methodologies to analyze these effects: (i) panel autoregressive distributed lag (ARDL) models and (ii) panel local projection models. Our results indicate that neither temperature nor precipitation anomalies have a statistically significant effect on GDP per capita, headline inflation, or their components. To the best of our knowledge, no previous study in Mexico has used temperature and precipitation anomalies to examine their effects on both GDP per capita and inflation, nor applied the methodological combination employed in this work. Finally, this work does not indicate that macroeconomic variables are unaffected by global warming and climate change, it indicates that temperature and precipitation anomalies registered in Mexico do not significantly affect either inflation or GDP per capita.  Future research should extend this analysis by incorporating extreme weather events, regional heterogeneity in adaptive capacity, and supply-chain linkages as additional potential drivers of inflation and output dynamics across Mexican regions.

\clearpage
\newpage
\bibliographystyle{apalike}  
\bibliography{Biblio}

@TECHREPORT{Andersson2020,
title = {Climate change and the macroeconomy},
author = {Andersson, Malin and Baccianti, Claudio and Morgan, Julian},
year = {2020},
institution = {European Central Bank},
type = {Occasional Paper Series},
number = {243},
url = {https://EconPapers.repec.org/RePEc:ecb:ecbops:2020243}
}

@article{Colacito2019,
author = {Colacito, Riccardo and Hoffmann, Bridget and Phan, Toan},
title = {Temperature and Growth: {A} Panel Analysis of the {United States}},
journal = {Journal of Money, Credit and Banking},
volume = {51},
number = {2-3},
pages = {313-368},
doi = {https://doi.org/10.1111/jmcb.12574},
url = {https://onlinelibrary.wiley.com/doi/abs/10.1111/jmcb.12574},
eprint = {https://onlinelibrary.wiley.com/doi/pdf/10.1111/jmcb.12574},
year = {2019}
}

@article{dell2012,
  title={Temperature shocks and economic growth: {E}vidence from the last half century},
  author={Dell, Melissa and Jones, Benjamin F and Olken, Benjamin A},
  journal={American Economic Journal: {M}acroeconomics},
  volume={4},
  number={3},
  pages={66--95},
  year={2012},
  publisher={American Economic Association}
}

@techreport{sudarshan2014economic,
  title={The economic impacts of temperature on industrial productivity: {E}vidence from {I}ndian manufacturing},
  author={Sudarshan, Anant and Tewari, Meenu},
  year={2014},
  type={Working paper},
  institution = {Indian Council for Research on International Economic Relations (ICRIER)},
  number = {278},
  address = {New Delhi},
}

@techreport{kotz2023impact,
    title={The impact of global warming on inflation: {A}verages, seasonality and extremes},
    author={Kotz, Maximilian and Kuik, Friderike and Lis, Eliza and Nickel, Christiane},
    year={2023},
    institution = {ECB},
    series  = {ECB Working Paper},
    type    = {Working Paper},
    number  = {2821}
}

@techreport{cevik2023eye,
  title={Eye of the storm: {T}he impact of climate shocks on inflation and growth},
  author={Cevik, Mr Serhan and Jalles, Jo{\~a}o Tovar},
  year={2023},
  institution={International Monetary Fund},
  type = {Working Paper},
  number = {WP/23/87}
}

@techreport{faccia2021feeling,
  title={Feeling the heat: {E}xtreme temperatures and price stability},
  author={Faccia, Donata and Parker, Miles and Stracca, Livio},
  year={2021},
  xmonth  = {December},
  institution={ECB Working Paper},
  type={Working Paper},
  number={2626},
}

@techreport{dafermos2021price,
    address = {Munich},
    author  = {Dafermos, Yannis and Kriwoluzky, Alexander and Vargas, Mauricio and Volz, Ulrich and Wittich, Jana},
    institution = {DIW Berlin: {P}olitikberatung kompakt},
    type    = {Working Paper},
    number  = {173},
    title   = {The price of hesitation: {H}ow the climate crisis threatens price stability and what the ECB must do about it},
    year    = {2021},
    xmonth  = {september},
}

@article{parker2018impact,
  title={The impact of disasters on inflation},
  author={Parker, Miles},
  journal={Economics of Disasters and Climate Change},
  volume={2},
  number={1},
  pages={21--48},
  year={2018},
  publisher={Springer}
}

@article{heinen2019price,
  title={The price impact of extreme weather in developing countries},
  author={Heinen, Andr{\'e}as and Khadan, Jeetendra and Strobl, Eric},
  journal={The Economic Journal},
  volume={129},
  number={619},
  pages={1327--1342},
  year={2019},
  publisher={Oxford University Press}
}

@techreport{batten2018climate,
  title={{Climate change and the macro-economy: {A} critical review}},
  author={Batten, Sandra},
  year={2018},
  publisher={Bank of England Working Paper},
type = {Working Papers},
number ={76},
institution = {Bank of England}
}

@article{acevedo2020effects,
  title={The effects of weather shocks on economic activity: {W}hat are the channels of impact?},
  author={Acevedo, Sebastian and Mrkaic, Mico and Novta, Natalija and Pugacheva, Evgenia and Topalova, Petia},
  journal={Journal of Macroeconomics},
  volume={65},
  pages={103207},
  year={2020},
  publisher={Elsevier}
}

@techreport{zhang2015temperature,
  title   = {Temperature and economic growth: {N}ew evidence from total factor productivity},
  author  = {Zhang, Peng},
  type    = {Working Paper},
  year    = {2015},
  xmonth  = {August},
 institution = {available at SSRN: dx.doi.org/10.2139/ssrn.2654406}, 
}

@article{letta2019weather,
  title={Weather, climate and total factor productivity},
  author={Letta, Marco and Tol, Richard SJ},
  journal={Environmental and Resource Economics},
  volume={73},
  number={1},
  pages={283--305},
  year={2019},
  publisher={Springer}
}

@article{alvi2021assessing,
  title={Assessing the impact of global warming on productivity in emerging economies of {A}sia},
  author={Alvi, Shahzad and Jamil, Faisal and Ahmed, Ather Maqsood},
  journal={International Journal of Global Warming},
  volume={24},
  number={2},
  pages={222--234},
  year={2021},
  publisher={Inderscience Publishers (IEL)}
}

@article{zhang2018temperature,
  title={Temperature effects on productivity and factor reallocation: {E}vidence from a half million Chinese manufacturing plants},
  author={Zhang, Peng and Deschenes, Olivier and Meng, Kyle and Zhang, Junjie},
  journal={Journal of Environmental Economics and Management},
  volume={88},
  pages={1--17},
  year={2018},
  publisher={Elsevier}
}

@article{jorda2005estimation,
  title={Estimation and inference of impulse responses by local projections},
  author={Jord{\`a}, {\`O}scar},
  journal={American Economic Review},
  volume={95},
  number={1},
  pages={161--182},
  year={2005},
  publisher={American Economic Association}
}

@article{kahn2021long,
  title={Long-term macroeconomic effects of climate change: {A} cross-country analysis},
  author={Kahn, Matthew E and Mohaddes, Kamiar and Ng, Ryan NC and Pesaran, M Hashem and Raissi, Mehdi and Yang, Jui-Chung},
  journal={Energy Economics},
  volume={104},
  pages={105624},
  year={2021},
  publisher={Elsevier}
}

@techreport{arellano2023temperature,
  title={{Temperature shocks and their effect on the price of agricultural products: {P}anel data evidence from vegetables in Mexico}},
  author={Arellano-Gonzalez, Jesus and Juárez-Torres, Miriam and Zazueta Borboa, Francisco},
  year={2023},
  institution={Bank of Mexico},
    type = {{Working Papers}},
number = {N° 2023-02}
}

@techreport{arellano2023weather_,
  title={{Weather shocks, prices and productivity: {E}vidence from staples in Mexico}},
  author={Arellano-Gonzalez, Jesus and Juárez-Torres, Miriam and Zazueta Borboa, Francisco},
  year={2023},
  institution={Bank of Mexico},
type ={Working Papers},
number ={N° 2023-16}
}

@article{arguez2012noaa,
  title={{NOAA's 1981--2010 {U.S.} climate normals: {A}n overview}},
  author={Arguez, Anthony and Durre, Imke and Applequist, Scott and Vose, Russell S and Squires, Michael F and Yin, Xungang and Heim, Richard R and Owen, Timothy W},
  journal={Bulletin of the American Meteorological Society},
  volume={93},
  number={11},
  pages={1687--1697},
  year={2012},
  publisher={American Meteorological Society}
}

@article{vose2014improved,
  title={Improved historical temperature and precipitation time series for {U.S.} climate divisions},
  author={Vose, Russell S and Applequist, Scott and Squires, Mike and Durre, Imke and Menne, Matthew J and Williams Jr, Claude N and Fenimore, Chris and Gleason, Karin and Arndt, Derek},
  journal={Journal of Applied Meteorology and Climatology},
  volume={53},
  number={5},
  pages={1232--1251},
  year={2014}
}

@article{DriscollKraay,
 ISSN = {00346535, 15309142},
 URL = {http://www.jstor.org/stable/2646837},
 author = {John C. Driscoll and Aart C. Kraay},
 journal = {The Review of Economics and Statistics},
 number = {4},
 pages = {549--560},
 publisher = {The MIT Press},
 title = {Consistent Covariance Matrix Estimation with Spatially Dependent Panel Data},
 urldate = {2025-03-28},
 volume = {80},
 year = {1998}
}

@techreport{akyapi2022estimating,
  title={Estimating macro-fiscal effects of climate shocks from billions of geospatial weather observations},
  author={Akyapi, Berkay and Bellon, Matthieu and Massetti, Emanuele},
  year={2022},
  institution={International Monetary Fund},
  type = {{Working Paper}},
  number = {WP/22/156}
}

@article{liu2025macroeconomic,
  title={Macroeconomic effects of climate change: {E}vidence from {C}anadian provinces},
  author={Liu, Lucy Q and Pan, Dan and Raissi, Mehdi},
  journal={International Economics},
  volume={181},
  pages={100572},
  year={2025},
  publisher={Elsevier}
}

@techreport{bundesbank2022,
title = {Climate change and climate policy: {A}nalytical requirements and options from a central bank perspective},
author = {{Deutsche Bundesbank}},
institution = {{Deutsche Bundesbank}},
type = {{Monthly Report - January}},
year = {2022},
pages = {33--62}
}

@techreport{ehlers2025,
  title={Macroeconomic impact of extreme weather events},
  author={Ehlers, Torsten and Frost, Jon and Madeira, Carlos and Shim, Ilhyock},
  year={2025},
  institution={Bank for International Settlements},
type = {{BIS bulletin No. 38}},
}

@misc{BNPParibas2022,
  author       = "{BNP Paribas}",
  title        = "Climate and inflation: {H}otter prices for a cooler planet",
  howpublished = "Podcast-October",
  year         = "2022",
  note         = "{BNP Paribas Global Markets’ --Strategy and Economics division}"
}

@article{plagborg2021,
  title={Local projections and {VAR}s estimate the same impulse responses},
  author={Plagborg-M{\o}ller, Mikkel and Wolf, Christian K},
  journal={Econometrica},
  volume={89},
  number={2},
  pages={955--980},
  year={2021},
  publisher={Wiley Online Library}
}

@techreport{olea2024double,
  title={Double robustness of local projections and some unpleasant {VAR}ithmetic},
  author={Olea, Jos{\'e} Luis Montiel and Plagborg-M{\o}ller, Mikkel and Qian, Eric and Wolf, Christian K},
  year={2024},
  institution={National Bureau of Economic Research},
  type = {Working Paper},
  series = {Working Paper Series},
  number = {32495} 
}

@article{tol2017population,
  title={Population and trends in the global mean temperature},
  author={Tol, Richard SJ},
  journal={Atm{\'o}sfera},
  volume={30},
  number={2},
  pages={121--135},
  year={2017},
  publisher={Centro de Ciencias de la Atm{\'o}sfera, UNAM}
}

@incollection{Baumeister2025,
  author  = {Baumeister, Christiane},
  title   = {{Discussion of \say{Local Projections or VARs? A Primer for Macroeconomists} by Jos{\'e} Luis Montiel Olea, Mikkel Plagborg-M{\"o}ller, Eric Qian, and Christian K. Wolf}},
  booktitle = {NBER Macroeconomics Annual 2025, volume 40},
  year    = {2025},
  publisher = {University of Chicago Press},
  editor = {Leahy, John V. and Ramey, Valerie A.}
}

@article{sims1980,
  author  = {Sims, Christopher A.},
  title   = {Macroeconomics and Reality},
  journal = {Econometrica},
  volume  = {48},
  number  = {1},
  pages   = {1--48},
  year    = {1980}
}

@article{smetswouters2003,
  author  = {Smets, Frank and Wouters, Rafael},
  title   = {An Estimated Dynamic Stochastic General Equilibrium Model of the Euro Area},
  journal = {Journal of the European Economic Association},
  volume  = {1},
  number  = {5},
  pages   = {1123--1175},
  year    = {2003}
}

@article{delnegroschorfheide2004,
  author  = {Del Negro, Marco and Schorfheide, Frank},
  title   = {Priors from General Equilibrium Models for {VAR}s},
  journal = {International Economic Review},
  volume  = {45},
  number  = {2},
  pages   = {643--673},
  year    = {2004}
}

@article{delnegroschorfheide2006,
  author  = {Del Negro, Marco and Schorfheide, Frank},
  title   = {How Good Is What You've Got? {DSGE-VAR} as a Toolkit for Evaluating {DSGE} Models},
  journal = {Economic Review, Federal Reserve Bank of Atlanta},
  volume  = {91},
  number  = {2},
  pages   = {21--37},
  year    = {2006}
}

@techreport{giacomini2013,
  author      = {Giacomini, Raffaella},
  title       = {The Relationship between {DSGE} and {VAR} Models},
  institution = {Centre for Microdata Methods and Practice (CeMMAP)},
  type        = {Working Paper},
  number      = {CWP21/13},
  year        = {2013}
}

@article{stockwatson2001,
  author  = {Stock, James H. and Watson, Mark W.},
  title   = {Vector Autoregressions},
  journal = {Journal of Economic Perspectives},
  volume  = {15},
  number  = {4},
  pages   = {101--115},
  year    = {2001}
}

@book{lutkepohl2005,
  author    = {L{\"u}tkepohl, Helmut},
  title     = {New Introduction to Multiple Time Series Analysis},
  publisher = {Springer},
  year      = {2005}
}

@techreport{Ludwig2024,
  author      = {Ludwig, Julian F.},
  title       = {{Local Projections are VAR predictions of increasing order}},
  institution = {Texas Tech University},
  type        = {Working Paper},
  year        = {2024}
}

@article{BurkeHsiangMiguel2015,
  author  = {Burke, Marshall and Hsiang, Solomon M. and Miguel, Edward},
  title   = {Global Non‐linear Effect of Temperature on Economic Production},
  journal = {Nature},
  volume  = {527},
  number  = {7577},
  pages   = {235--239},
  year    = {2015},
  doi     = {10.1038/nature15725}
}

@article{LiCongGuZhang2021,
  author  = {Li, Chengzheng and Cong, Jiajia and Gu, Haiying and Zhang, Peng},
  title   = {The Non‐linear Effect of Daily Weather on Economic Performance: Evidence from {C}hina},
  journal = {China Economic Review},
  volume  = {69},
  pages   = {101647},
  year    = {2021},
  doi     = {10.1016/j.chieco.2021.101647}
}

@article{DellJonesOlken2012,
  author  = {Dell, Melissa and Jones, Benjamin F. and Olken, Benjamin A.},
  title   = {Temperature Shocks and Economic Growth: Evidence from the Last Half Century},
  journal = {American Economic Journal: Macroeconomics},
  volume  = {4},
  number  = {3},
  pages   = {66--95},
  year    = {2012},
  doi     = {10.1257/mac.4.3.66}
}

@article{ZhangDeschenesMeng2017,
  author  = {Zhang, Peng and Desch\^enes, Olivier and Meng, Kyle C. and Zhang, Junjie},
  title   = {Temperature Effects on Productivity and Factor Reallocation: Evidence from a Half Million Chinese Manufacturing Plants},
  journal = {Journal of Environmental Economics and Management},
  volume  = {88},
  pages   = {1--17},
  year    = {2018},
  doi     = {10.1016/j.jeem.2017.11.003}
}

@misc{DongTolWang2025,
  author       = {Dong, Jinchi and Tol, Richard S.~J. and Wang, Jinnan},
  title        = {The Effects of Climate and Weather on Economic Output: Evidence from Global Subnational Data},
  howpublished = {arXiv Working Paper},
  year         = {2025},
  note         = {Available at https://arxiv.org/abs/2505.17946}
}

@article{binder1999stochastic,
  title={Stochastic growth models and their econometric implications},
  author={Binder, Michael and Pesaran, M Hashem},
  journal={Journal of Economic Growth},
  volume={4},
  number={2},
  pages={139--183},
  year={1999},
  publisher={Springer}
}

@article{pesaran2001bounds,
  title={Bounds testing approaches to the analysis of level relationships},
  author={Pesaran, M Hashem and Shin, Yongcheol and Smith, Richard J},
  journal={{Journal of Applied Econometrics}},
  volume={16},
  number={3},
  pages={289--326},
  year={2001},
  publisher={Wiley Online Library}
}

@article{cho2023recent,
  title={Recent developments of the autoregressive distributed lag modelling framework},
  author={Cho, Jin Seo and Greenwood-Nimmo, Matthew and Shin, Yongcheol},
  journal={Journal of Economic Surveys},
  volume={37},
  number={1},
  pages={7--32},
  year={2023},
  publisher={Wiley Online Library}
}

@incollection{Pesaran_Shin_1999, 
    place={Cambridge}, 
    series={Econometric Society Monographs}, 
    title={An Autoregressive Distributed-Lag Modelling Approach to Cointegration Analysis}, 
    booktitle={Econometrics and Economic Theory in the 20th Century: The Ragnar Frisch Centennial Symposium}, 
    publisher={Cambridge University Press}, 
    author={Pesaran, M. Hashem and Shin, Yongcheol}, 
    editor={Strøm, SteinarEditor}, 
    year={1999}, 
    pages={371–413}, 
    collection={Econometric Society Monographs}
}

@book{pesaran2015time,
  title={Time series and panel data econometrics},
  author={Pesaran, M Hashem},
  year={2015},
  publisher={Oxford University Press}
}

@article{mendelsohn2010ricardian,
  title={A {R}icardian analysis of {M}exican farms},
  author={Mendelsohn, Robert and Arellano-Gonzalez, Jesus and Christensen, Peter},
  journal={Environment and Development Economics},
  volume={15},
  number={2},
  pages={153--171},
  year={2010},
  publisher={Cambridge University Press}
}

@article{arellano2025temperature,
  title={Temperature and quarterly economic activity: {P}anel data evidence from {M}exico},
  author={Arellano-Gonz{\'a}lez, Jes{\'u}s and Ju{\'a}rez-Torres, Miriam},
  journal={Environment and Development Economics},
  pages={1--21},
  year={2025},
  publisher={Cambridge University Press}
}

@misc{ibarran2025,
  author       = {Ibarrar\'an, Mar\'ia Eugenia and Estrada, Francisco and P\'erez, Tamara},
  title        = {La Econom\'ia del Cambio Clim\'atico en {M}\'exico 2.0},
  howpublished = {Working Paper},
  year         = {2025},
  note         = {Available at https://www.pincc.unam.mx/wp-content/uploads/2025/03/Eugenia-Ibarraran-2022-Economia-del-cambio-climatico-en-Mexico.pdf}
}

@article{estrada2022impacts,
  title={Impacts and economic costs of climate change on Mexican agriculture},
  author={Estrada, Francisco and Mendoza-Ponce, Alma and Calder{\'o}n-Bustamante, Oscar and Botzen, Wouter},
  journal={Regional Environmental Change},
  volume={22},
  number={4},
  pages={126},
  year={2022},
  publisher={Springer}
}

@techreport{kabundi2022persistent,
  title={How persistent are climate-related price shocks},
  author={Kabundi, Alain and Mlachila, Montfort and Yao, Jiaxiong},
  year={2022},
  series={IMF Working Papers},
  type={Working paper},
  institution = {International Monetary Fund},
  number = {22/207}
}

@techreport{costa2025macroeconomic,
  title={The macroeconomic implications of extreme weather events},
  author={Costa, H{\'e}lia and Hooley, John},
  institution={OECD},
  series={OECD Economics Department Working Papers},
  year={2025},
  type={Working paper},
  number={1837}
}

@article{ciccarelli2024demand,
  title={Demand or supply? {A}n empirical exploration of the effects of climate change on the macroeconomy},
  author={Ciccarelli, Matteo and Marotta, Fulvia},
  journal={Energy Economics},
  volume={129},
  pages={107163},
  year={2024},
  publisher={Elsevier}
}

@techreport{martinez2025macroeconomic,
  author      = {Andrew B. Martinez},
  title       = {How do Macroeconomic Expectations React to Extreme Weather Shocks?},
  institution = {U.S. Department of the Treasury, Office of Macroeconomic Analysis},
  number        = {2025-001},
  year= {2025},
  type        = {Working Paper}
}

@techreport{bilal2024macroeconomic,
  title={The macroeconomic impact of climate change: Global vs. local temperature},
  author={Bilal, Adrien and K{\"a}nzig, Diego R},
  year={2024},
  series={NBER Working Papers},
  institution={NBER},
  type={Working paper},
  number={32450}
}

@techreport{winter2023long,
  title={Long-term macroeconomic effects of shifting temperature anomaly distributions},
  author={Winter, David J and Kiehl, Manuela},
  series={SSRN Working Paper},
  number={4575937},
  year={2023},
  institution={SSRN},
}

@article{basurto2023impactos,
  title={Impactos econ{\'o}micos potenciales del cambio clim{\'a}tico en la ganader{\'\i}a: caso de {M}{\'e}xico},
  author={Basurto Hern{\'a}ndez, Sa{\'u}l and Galindo Paliza, Luis Miguel and R{\'\i}os Mohar, Julia},
  journal={Problemas del Desarrollo},
  volume={54},
  number={212},
  pages={27--54},
  year={2023},
  publisher={Universidad Nacional Aut{\'o}noma de M{\'e}xico, Instituto de Investigaciones Econ{\'o}micas}
}

@article{gay2006potential,
  title={Potential impacts of climate change on agriculture: {A} case of study of coffee production in {V}eracruz, {M}exico},
  author={Gay, Carlos and Estrada, Francisco and Conde, Cecilia and Eakin, Hallie and Villers, Lourdes},
  journal={Climatic Change},
  volume={79},
  number={3},
  pages={259--288},
  year={2006},
  publisher={Springer}
}

@article{estrada2023model,
  title={Model emulators for the assessment of regional impacts and risks of climate change: A case study of rainfed maize production in Mexico},
  author={Estrada, Francisco and Mendoza, Alma and Murray, Guillermo and Calder{\'o}n-Bustamante, Oscar and Botzen, Wouter and De Le{\'o}n Escobedo, Teresa and Velasco, Juli{\'a}n A},
  journal={Frontiers in Environmental Science},
  volume={11},
  pages={1027545},
  year={2023},
  publisher={Frontiers Media SA}
}


\setcounter{figure}{0}
\renewcommand{\thefigure}{A.\arabic{figure}}
\setcounter{table}{0}
\renewcommand{\thetable}{A.\arabic{table}}
\newpage

\appendix
\section{Additional descriptive results}

\begin{figure}[!htb]
    \centering
    \includegraphics[scale=1.1]{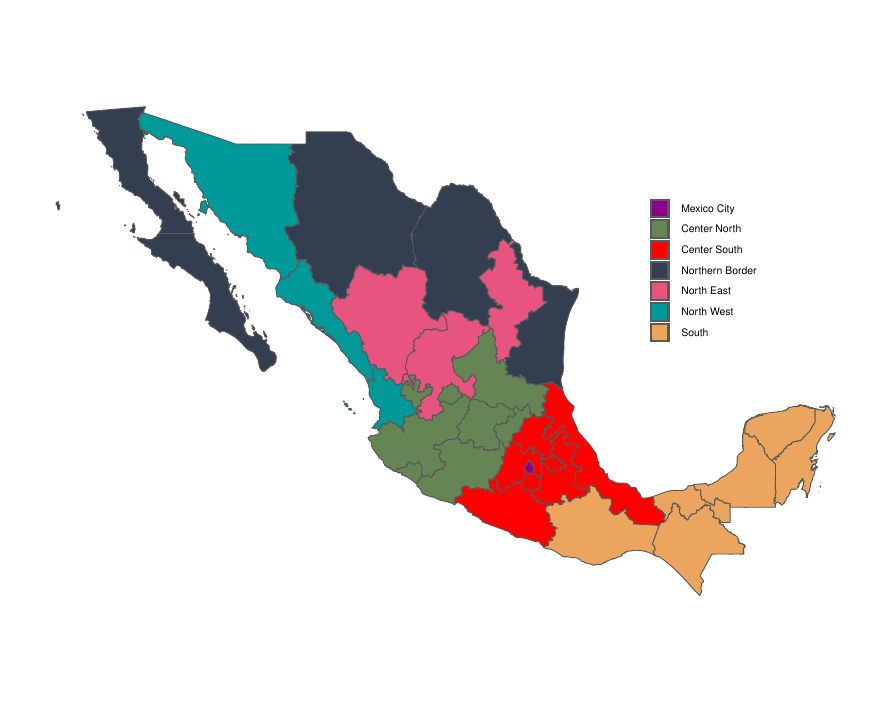}
    \caption{Regions in Mexico}
    \label{fig:map_regions}
\end{figure}

\begin{figure}[!htb]
    \centering
     \subfloat[(a) Headline inflation]{\includegraphics[scale=0.55]{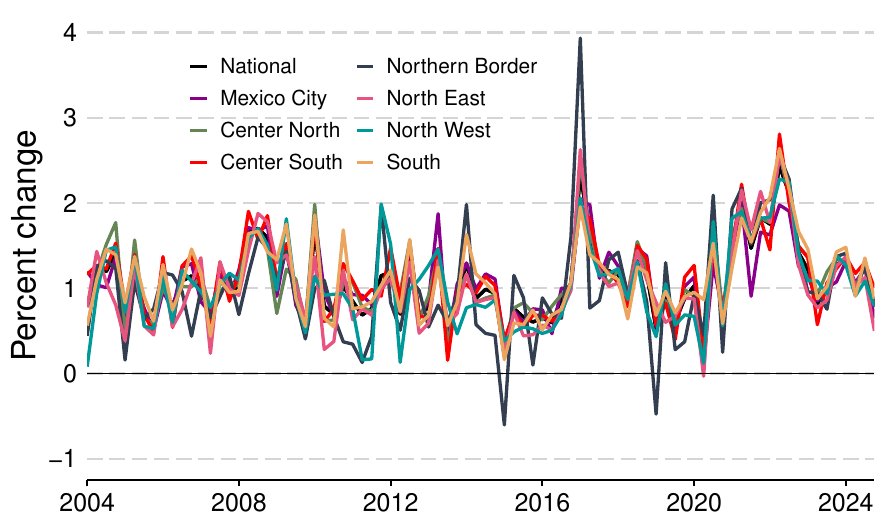}}
    \subfloat[(b) Food, beverages, and tobacco inflation]{\includegraphics[scale=0.55]{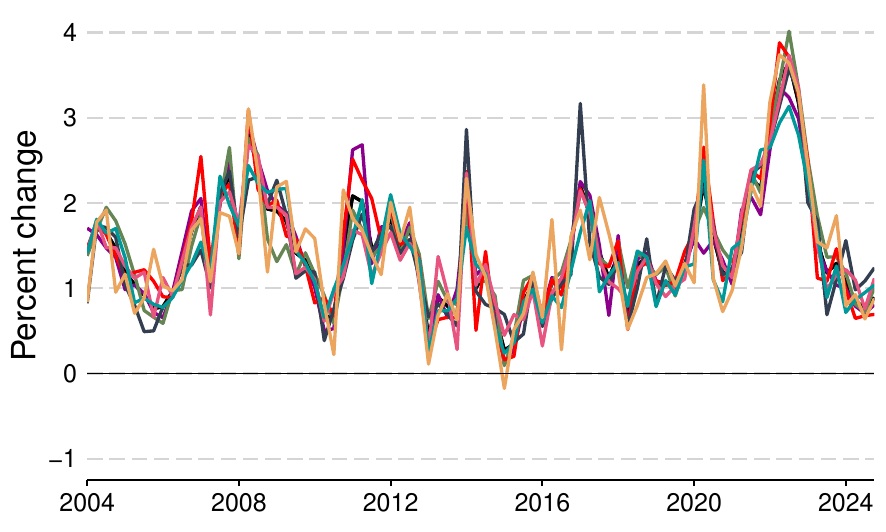}}
    \hfil
    \subfloat[(c) Non-food goods inflation]{\includegraphics[scale=0.55]{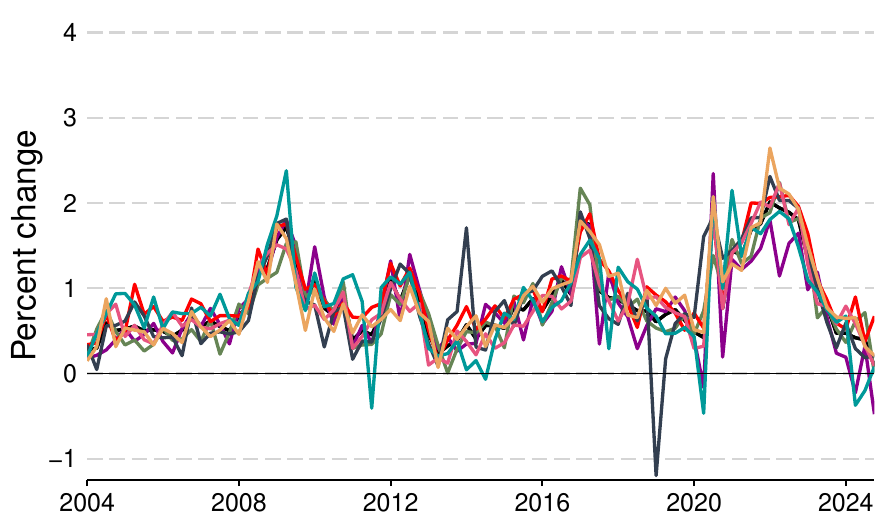}}
    \hfil
    \subfloat[(d) Services inflation]{\includegraphics[scale=0.55]{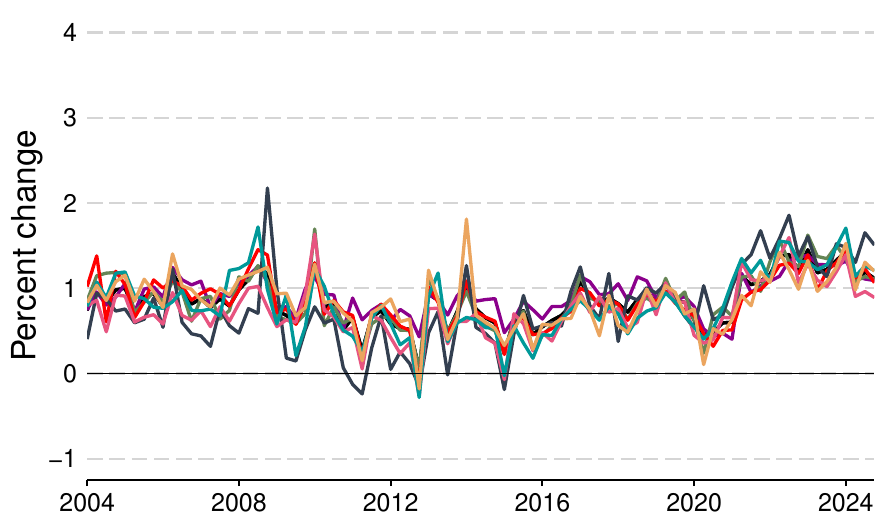}}
    \hfil
    \subfloat[(e) Agricultural inflation]{\includegraphics[scale=0.55]{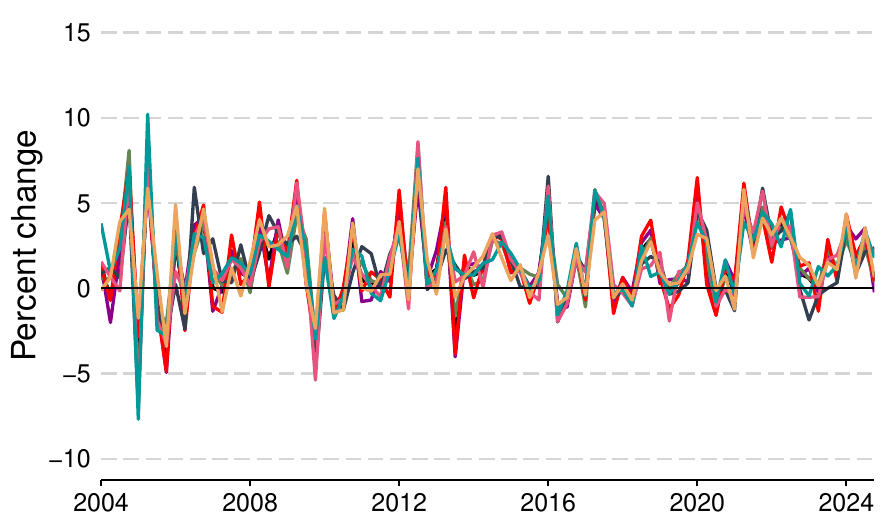}}
    \hfil
    \subfloat[(f) Energy inflation]{\includegraphics[scale=0.55]{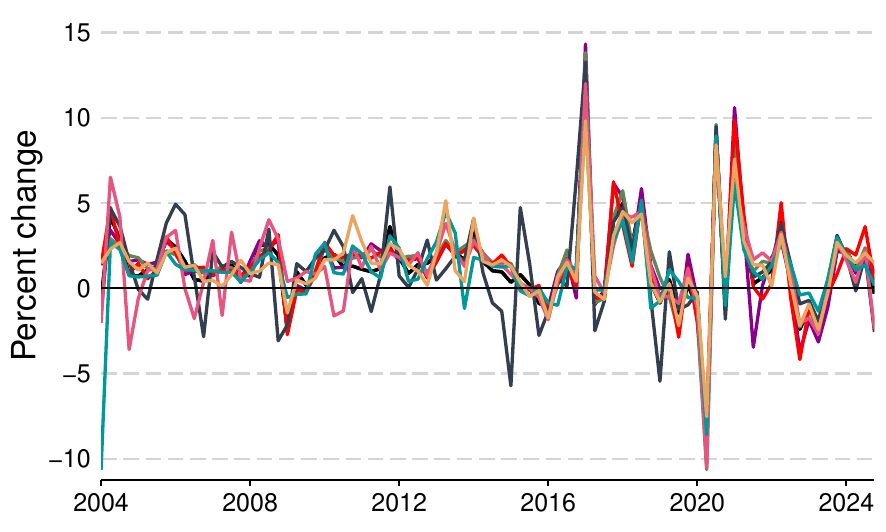}}
    \caption{Headline and selected CPI components quarterly inflation}
    \caption*{\small{\textbf{Source:} INEGI.\\ \textbf{Notes:} Inflation is calculated as the quarterly change of the seasonally adjusted Consumer Price Index.}}
    \vspace{-0.15cm}
    \label{fig:Region_sector_inf}
\end{figure}

\begin{figure}[!htb]
    \centering
    \subfloat[(a) Total real GDP per capita]{\includegraphics[scale=0.55]{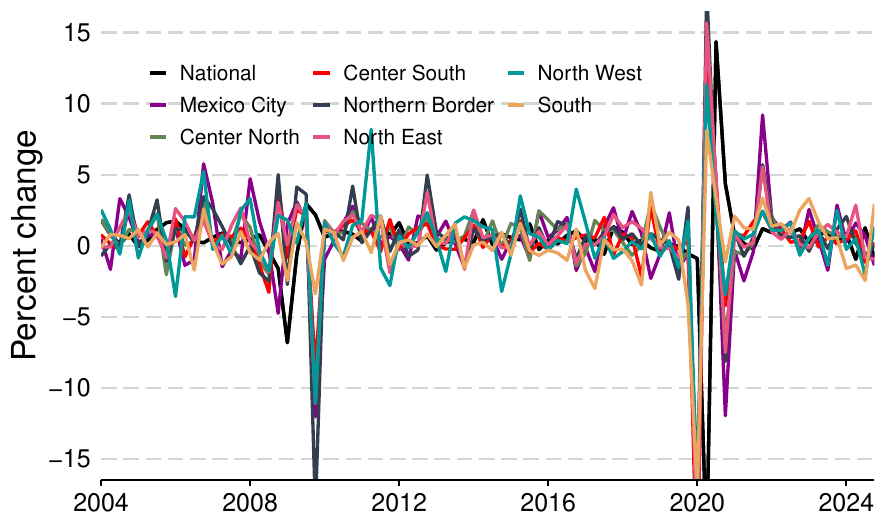}}
    \subfloat[(b) Primary sector real GDP per capita]{\includegraphics[scale=0.55]{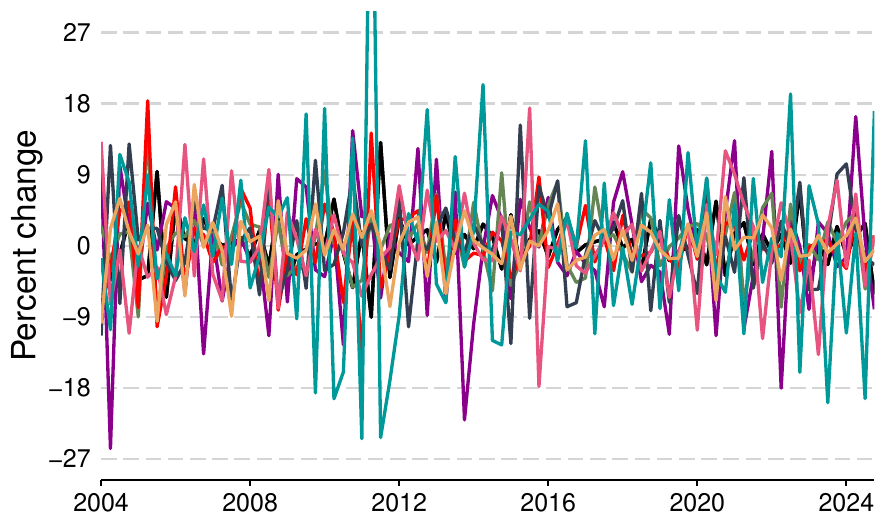}}
    \hfil
    \subfloat[(c) Secondary sector real GDP per capita]{\includegraphics[scale=0.55]{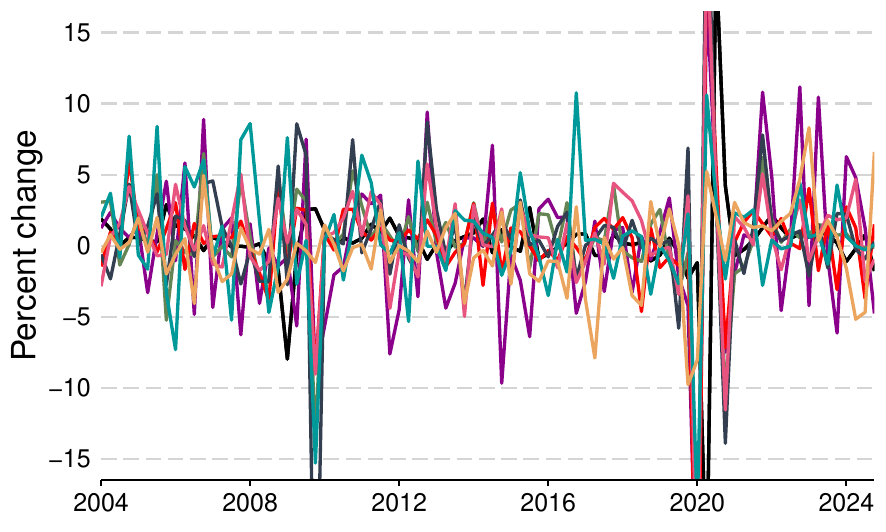}}
    \hfil
    \subfloat[(d) Tertiary sector real GDP per capita]{\includegraphics[scale=0.55]{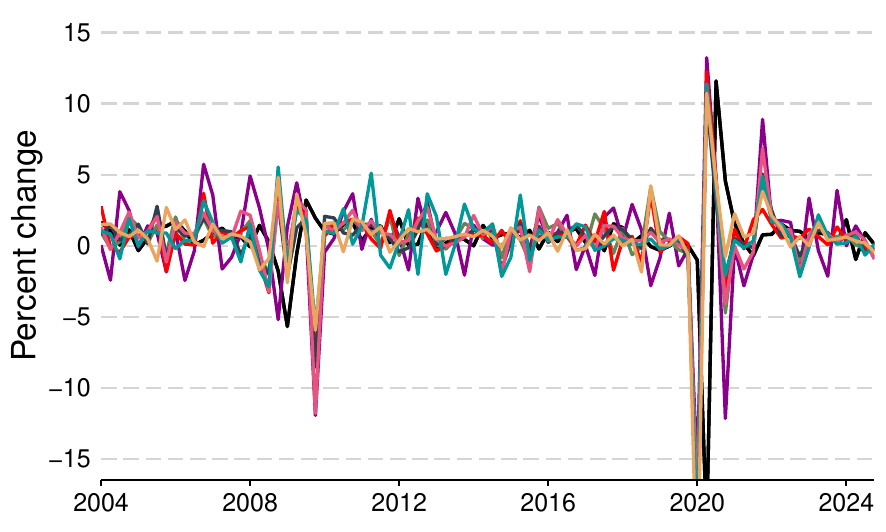}}
    \caption{Total and sectorial quarterly real GDP per capita}
    \caption*{\small{\textbf{Source:} INEGI.\\ \textbf{Notes:} The figure shows the seasonally adjusted real GDP per capita measured using 2018 Mexican pesos and the 2020 population level.}}
    \vspace{-0.15cm}
    \label{fig:Region_sector_gdp}
\end{figure}

\begin{figure}[htb!]
    \centering
    \subfloat[Temperature]{\includegraphics[scale=0.55]{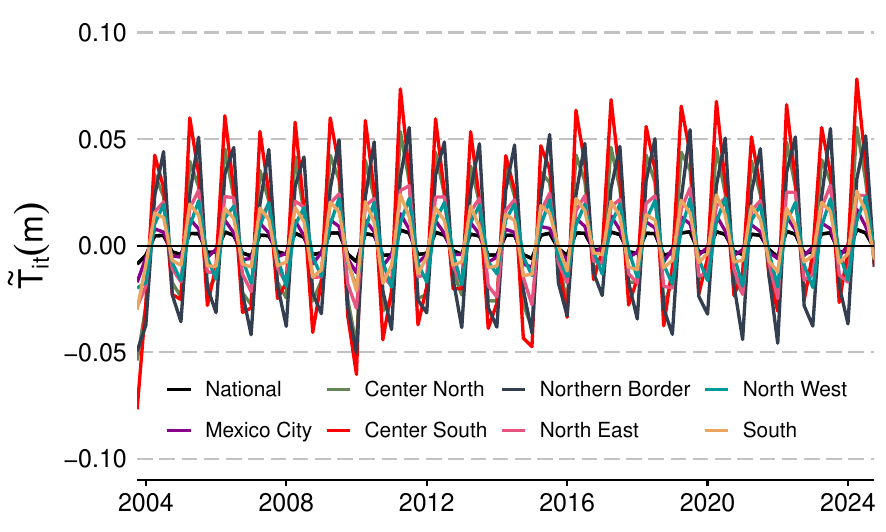}}
    \hfill
    \subfloat[Precipitation]{\includegraphics[scale=0.55]{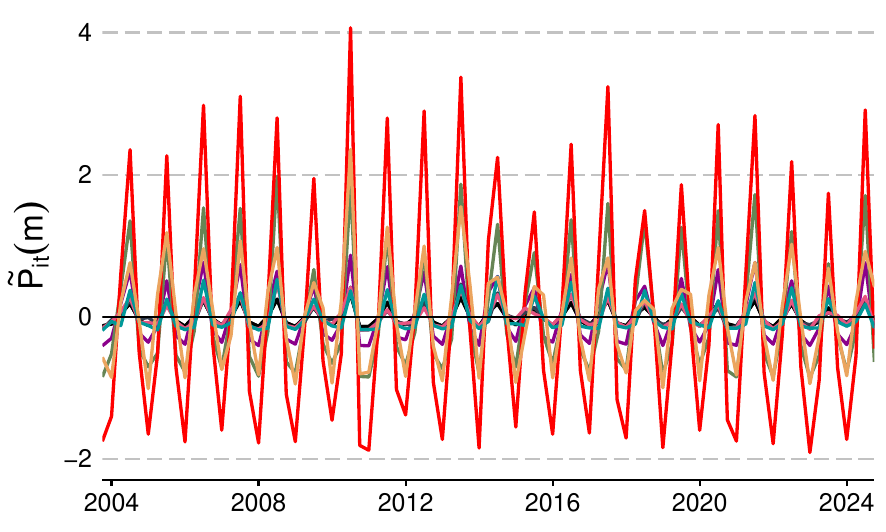}}
    \begin{center}
         \caption{Historical average of population-weighted temperature and precipitation anomalies.}\label{fig:TS_climate_reg}
        \vspace{-0.15cm}
        \caption*{\small{\textbf{Source:} World Bank.\\ \textbf{Notes:} Temperature and precipitation anomalies are computed as the deviation from their (30-year) norm. See \autoref{eq:temp_an}.}}
    \end{center}
\end{figure}

\begin{figure}
    \centering
    \subfloat[National]{\includegraphics[scale=0.475]{Descriptive/Climate/Temp-Distribution/Dist-TemperatureDev_7R-NATIONAL_Quarterly_MA30.pdf}}
    \hfil
    \subfloat[Northern-Border]{\includegraphics[scale=0.475]{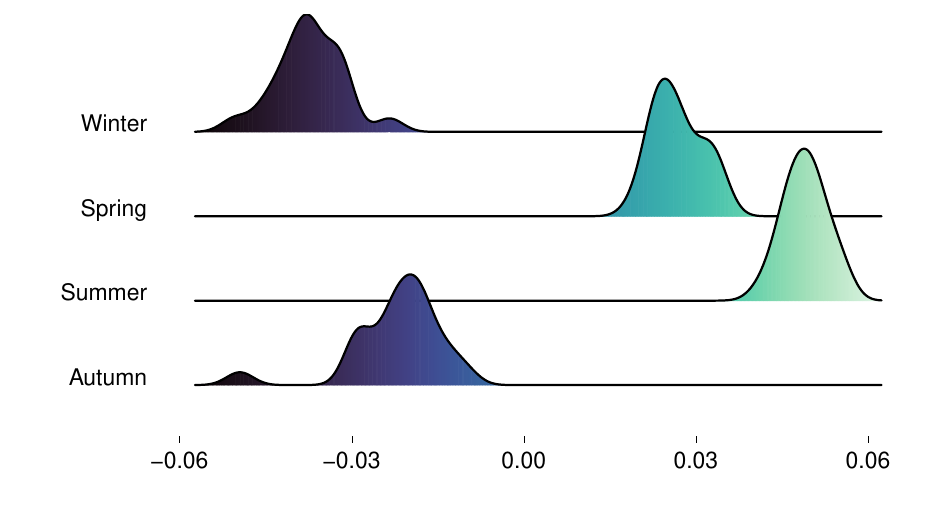}}
    \hfil
    \subfloat[North West]{\includegraphics[scale=0.475]{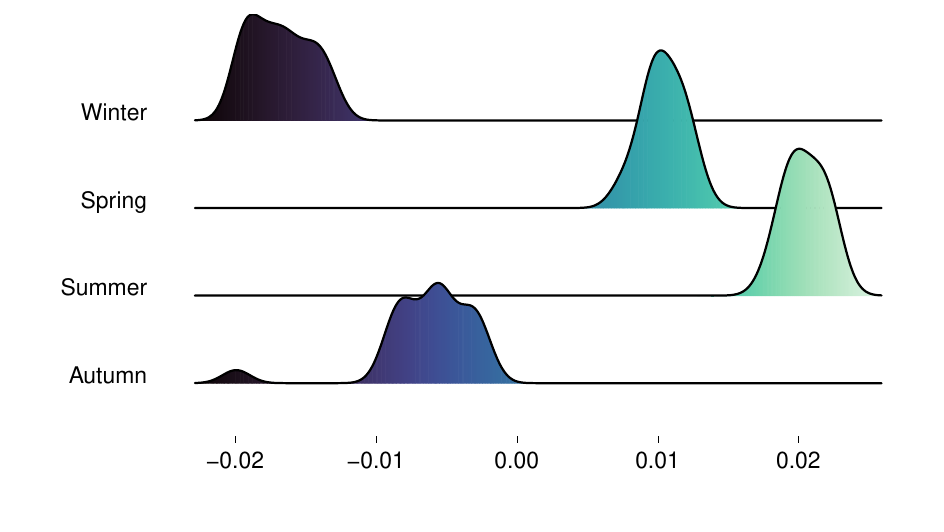}}
    \hfil
    \subfloat[North East]{\includegraphics[scale=0.475]{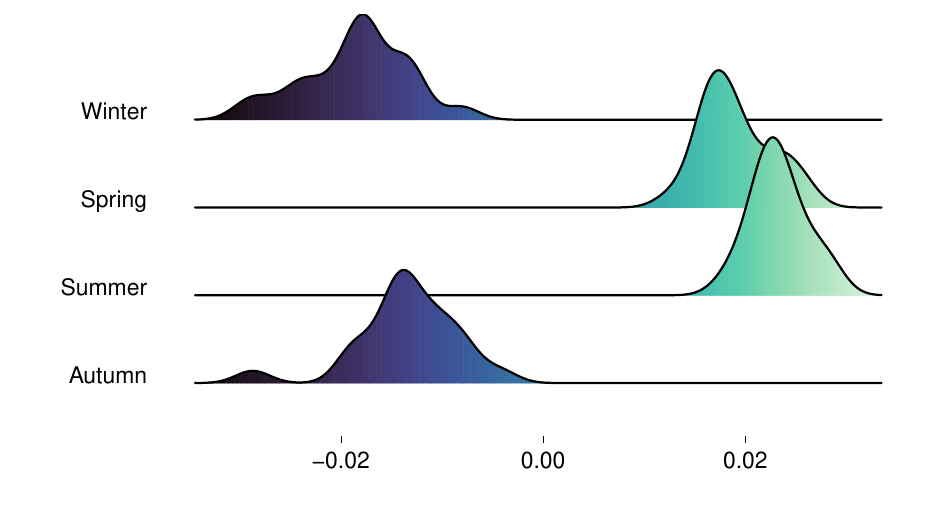}}
    \hfil
    \subfloat[Center North]{\includegraphics[scale=0.475]{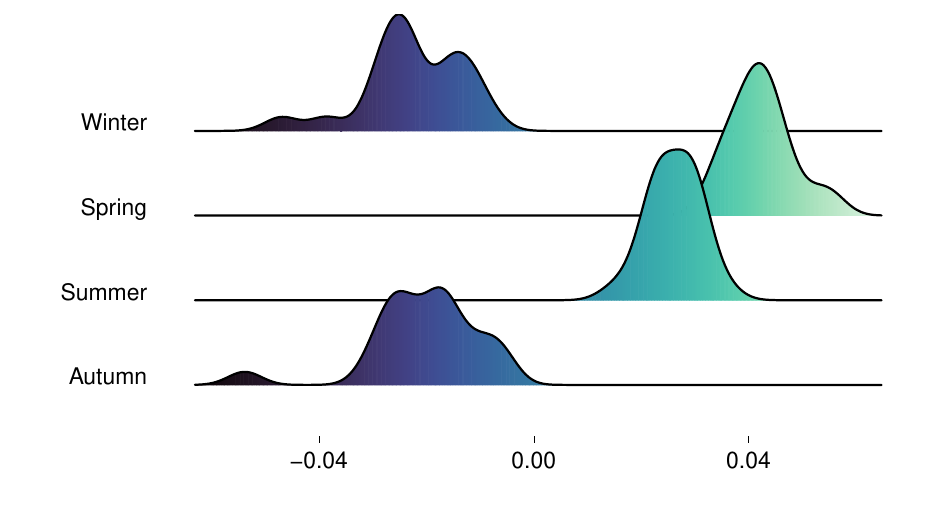}}
    \hfil
    \subfloat[Mexico City Metro. Area]{\includegraphics[scale=0.475]{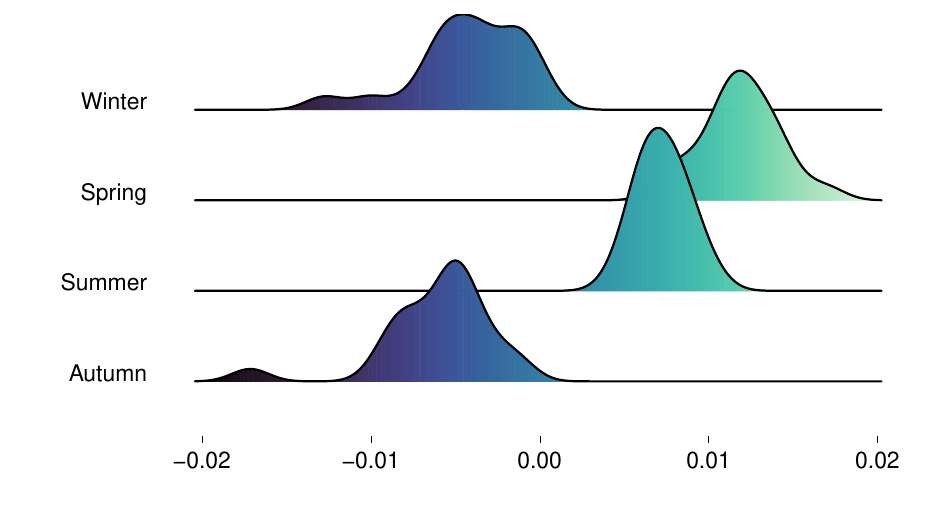}}
    \hfil
    \subfloat[Center South]{\includegraphics[scale=0.475]{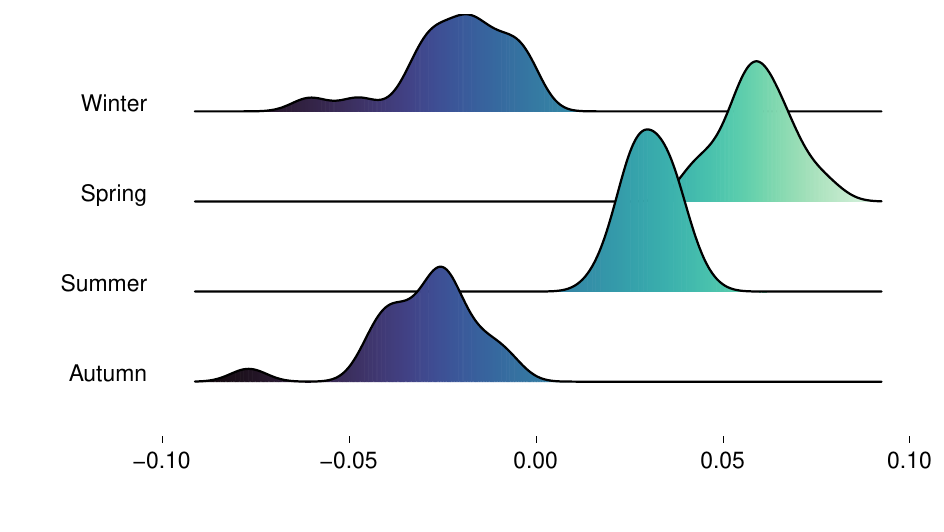}}
    \hfil
    \subfloat[South]{\includegraphics[scale=0.475]{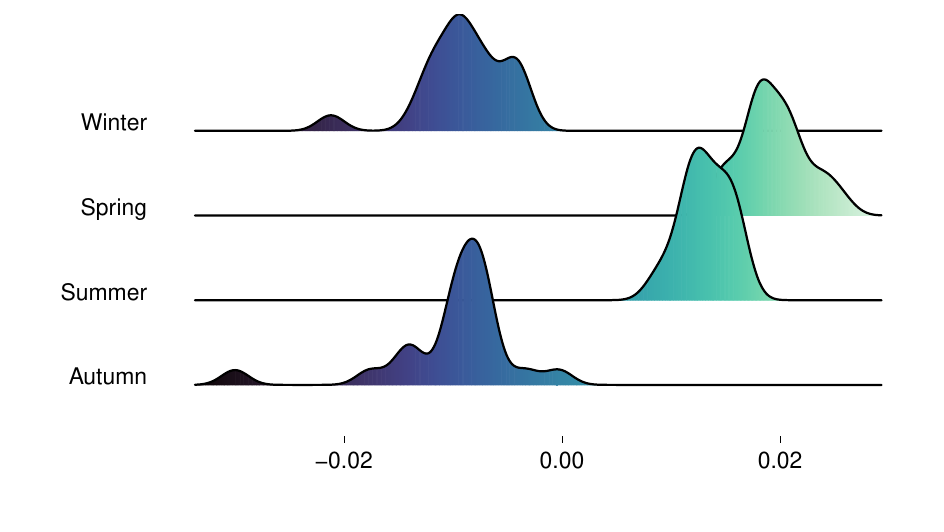}}
        \caption{Regional Population-weighted temperature anomalies distribution by season.} \label{fig:Regions_temp}
        \vspace{-0.15cm}
        \caption*{\footnotesize{\textbf{Source:} INEGI.\\ \textbf{Notes:} Temperature anomalies computed as deviations from their (30-year) norm. See \autoref{eq:temp_an}.}}
\end{figure}

\begin{figure}
    \centering
    \subfloat[National]{\includegraphics[scale=0.475]{Descriptive/Climate/Precip-Distribution/Dist-PrecipitationDev_7R-NATIONAL_Quarterly_MA30.pdf}}
    \hfil
    \subfloat[Northern-Border]{\includegraphics[scale=0.475]{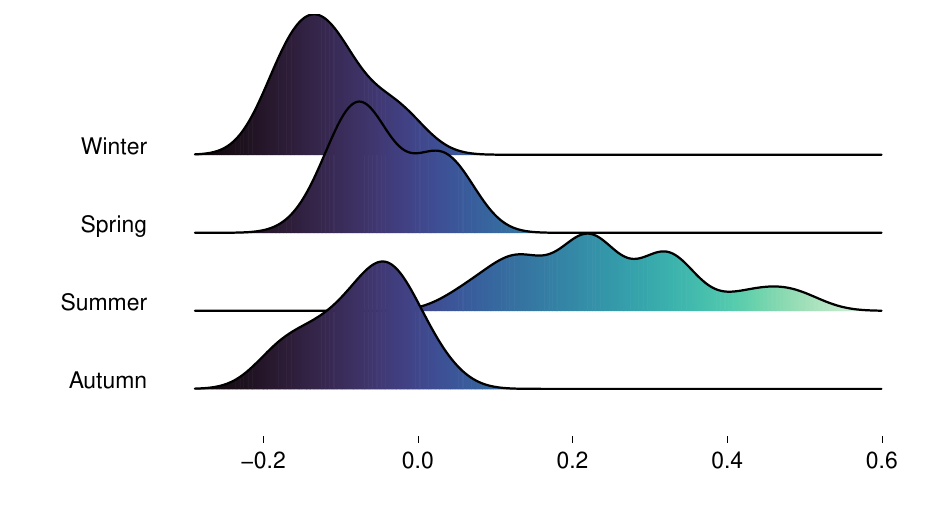}}
    \hfil
    \subfloat[North West]{\includegraphics[scale=0.475]{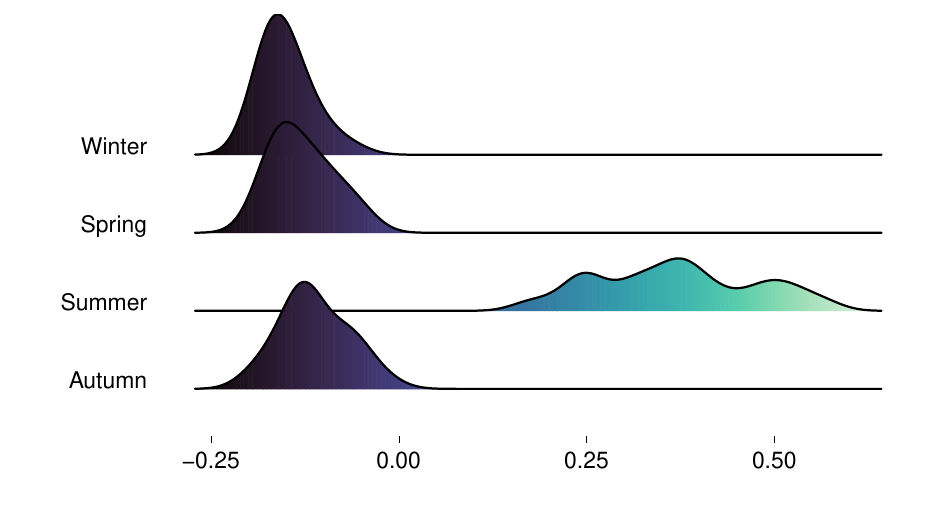}}
    \hfil
    \subfloat[North East]{\includegraphics[scale=0.475]{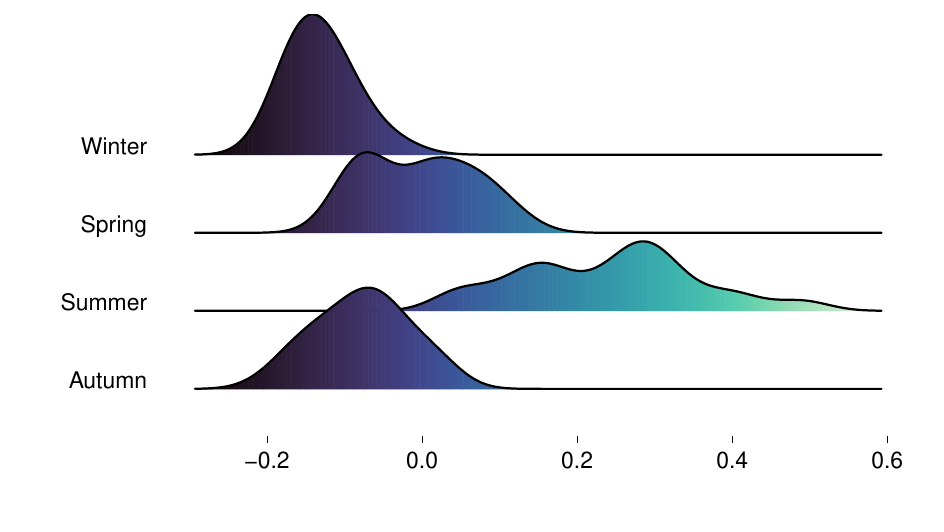}}
    \hfil
    \subfloat[Center North]{\includegraphics[scale=0.475]{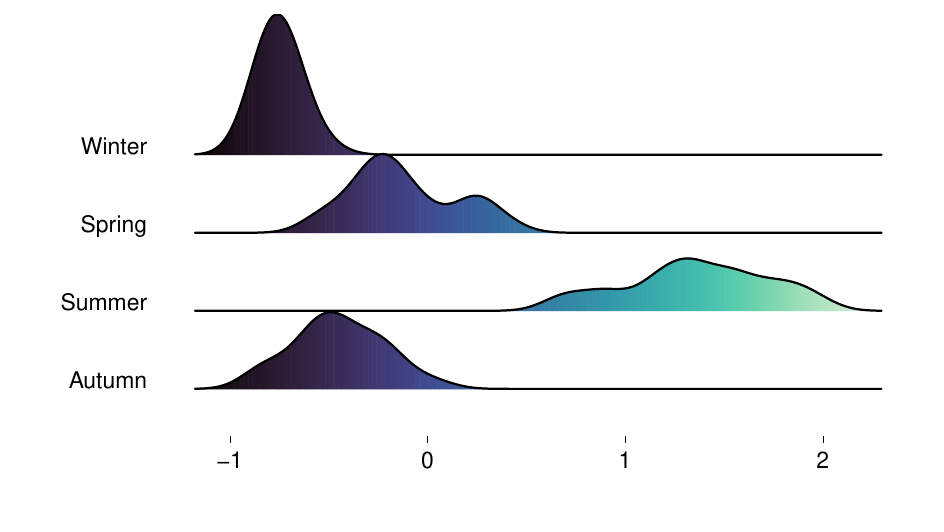}}
    \hfil
    \subfloat[Mexico City Metro. Area]{\includegraphics[scale=0.475]{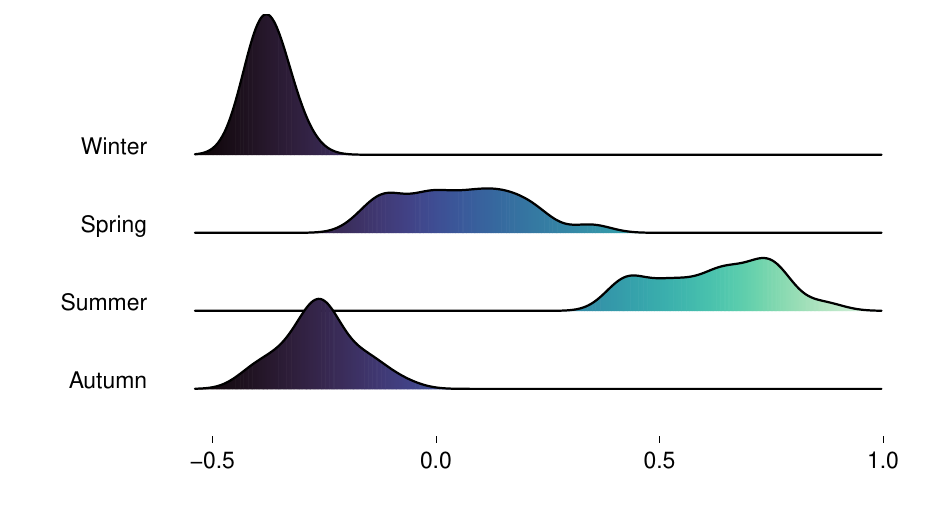}}
    \hfil
    \subfloat[Center South]{\includegraphics[scale=0.475]{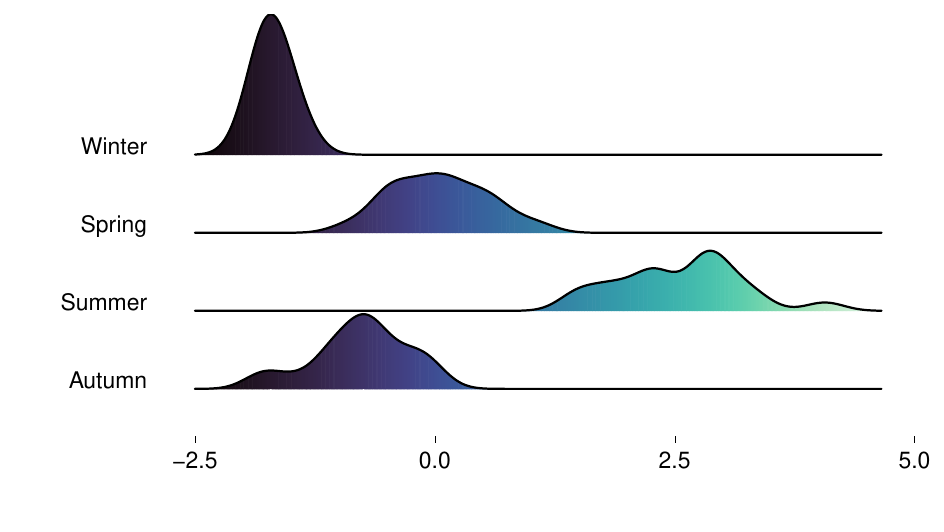}}
    \hfil
    \subfloat[South]{\includegraphics[scale=0.475]{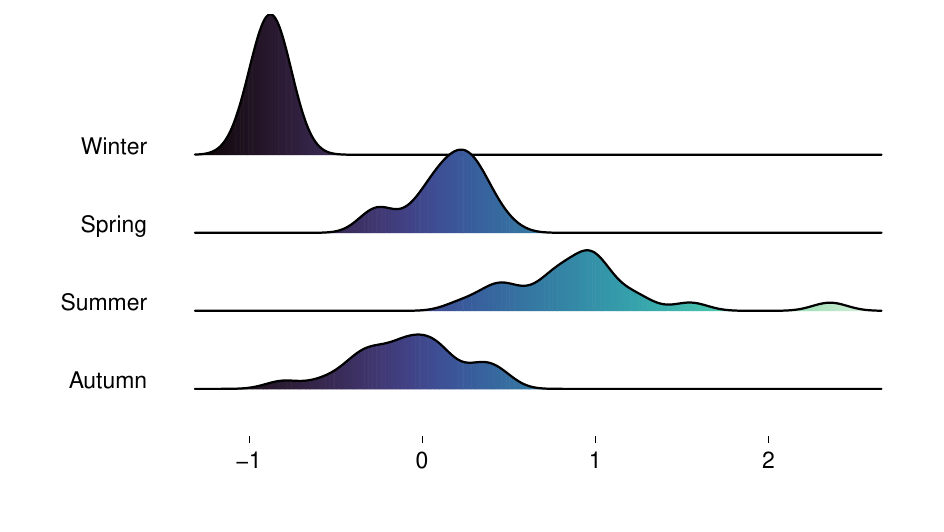}}
    \caption{Regional population-weighted precipitation anomalies distribution by season.} \label{fig:Regions_precip}
    \vspace{-0.15cm}
    \caption*{\footnotesize{\textbf{Source:} INEGI.\\ \textbf{Notes:} Precipitation anomalies computed as deviations from their (30-year) norm. See \autoref{eq:temp_an}.}}
\end{figure}


\clearpage
\section{Additional auxiliary results}

\begin{table}[!htb]
    \centering
    \caption{\small Long-run effects of positive and negative weather anomalies on GDP per capita by economic sectors.}
    \label{tab:sectors_ARDL}
    \vspace{0.15cm}
    \scriptsize 
    \begin{tabularx}{\textwidth}{mmmmmmmmmmmm}
         \toprule
         & & \multicolumn{3}{c}{Specification 1} & \multicolumn{3}{c}{Specification 2} & \multicolumn{3}{c}{Specification 3} \\
         \cmidrule(lr){3-5} \cmidrule(lr){6-8} \cmidrule(lr){9-11}
          &  & m = 20 & m = 30 & m = 40 & m = 20 & m = 30 & m = 40 & m = 20 & m = 30 & m = 40 \\ 
  \midrule
\multirow{10}{4em}{Primary} & $\hat{\theta}_{\Delta\Tilde{T}_{it}(m)^{+}}$ & -2.457 ** & -3.734 ** & -4.782 ** & -1.813  & -2.742  & -3.463  &  &  &  \\ 
   &  & (1.113) & (1.694) & (2.251) & (1.113) & (1.670) & (2.215) &  &  &  \\ 
   & $\hat{\theta}_{\Delta\Tilde{T}_{it}(m)^{-}}$ & -2.095 *** & -3.157 *** & -4.179 *** & -1.863 ** & -2.779 ** & -3.669 ** &  &  &  \\ 
   &  & (0.891) & (1.320) & (1.760) & (0.846) & (1.247) & (1.651) &  &  &  \\ 
   & $\hat{\theta}_{\Delta\Tilde{P}_{it}(m)^{+}}$ & -0.004  & -0.003  & -0.006  &  &  &  & -0.000  & 0.000  & -0.001  \\ 
   &  & (0.017) & (0.024) & (0.032) &  &  &  & (0.016) & (0.024) & (0.032) \\ 
   & $\hat{\theta}_{\Delta\Tilde{P}_{it}(m)^{-}}$ & -0.014  & -0.021  & -0.027  &  &  &  & -0.005  & -0.005  & -0.008  \\ 
   &  & (0.028) & (0.043) & (0.056) &  &  &  & (0.038) & (0.056) & (0.074) \\ 
   & $\hat{\phi}_{Primary}$ & 2.719 *** & 2.720 *** & 2.719 *** & 2.725 *** & 2.725 *** & 2.724 *** & 2.667 *** & 2.667 *** & 2.666 *** \\ 
   &  & (0.287) & (0.287) & (0.288) & (0.287) & (0.287) & (0.287) & (0.281) & (0.281) & (0.281) \\ 
   \midrule
\multirow{10}{4em}{Secondary} & $\hat{\theta}_{\Delta\Tilde{T}_{it}(m)^{+}}$ & -3.956  & -5.904  & -7.377  & -3.256  & -4.812  & -5.891  &  &  &  \\ 
   &  & (2.889) & (4.405) & (5.774) & (2.097) & (3.181) & (4.146) &  &  &  \\ 
   & $\hat{\theta}_{\Delta\Tilde{T}_{it}(m)^{-}}$ & -1.317  & -2.002  & -2.554  & -1.413  & -2.210  & -2.798  &  &  &  \\ 
   &  & (1.741) & (2.588) & (3.466) & (1.527) & (2.257) & (3.040) &  &  &  \\ 
   & $\hat{\theta}_{\Delta\Tilde{P}_{it}(m)^{+}}$ & 0.044  & 0.067  & 0.086  &  &  &  & 0.044 ** & 0.067 ** & 0.085 ** \\ 
   &  & (0.031) & (0.046) & (0.061) &  &  &  & (0.022) & (0.033) & (0.043) \\ 
   & $\hat{\theta}_{\Delta\Tilde{P}_{it}(m)^{-}}$ & 0.085  & 0.121  & 0.163  &  &  &  & 0.096  & 0.139  & 0.187  \\ 
   &  & (0.101) & (0.153) & (0.204) &  &  &  & (0.101) & (0.151) & (0.202) \\ 
   & $\hat{\phi}_{Secondary}$ & 1.639 *** & 1.635 *** & 1.634 *** & 1.692 *** & 1.693 *** & 1.695 *** & 1.654 *** & 1.653 *** & 1.653 *** \\ 
   &  & (0.326) & (0.325) & (0.325) & (0.349) & (0.352) & (0.354) & (0.348) & (0.348) & (0.348) \\ 
   \midrule
\multirow{10}{4em}{Tertiary} & $\hat{\theta}_{\Delta\Tilde{T}_{it}(m)^{+}}$ & -2.333  & -3.465  & -4.408  & -1.363  & -2.071  & -2.511  &  &  &  \\ 
   &  & (1.642) & (2.462) & (3.227) & (1.087) & (1.661) & (2.173) &  &  &  \\ 
   & $\hat{\theta}_{\Delta\Tilde{T}_{it}(m)^{-}}$ & -0.441  & -0.697  & -0.849  & -0.368  & -0.626  & -0.750  &  &  &  \\ 
   &  & (0.997) & (1.464) & (1.961) & (0.918) & (1.317) & (1.766) &  &  &  \\ 
   & $\hat{\theta}_{\Delta\Tilde{P}_{it}(m)^{+}}$ & 0.009  & 0.014  & 0.017  &  &  &  & 0.021  & 0.032  & 0.041  \\ 
   &  & (0.025) & (0.036) & (0.047) &  &  &  & (0.018) & (0.027) & (0.035) \\ 
   & $\hat{\theta}_{\Delta\Tilde{P}_{it}(m)^{-}}$ & 0.060  & 0.085  & 0.116  &  &  &  & 0.046  & 0.064  & 0.090  \\ 
   &  & (0.073) & (0.108) & (0.144) &  &  &  & (0.064) & (0.094) & (0.125) \\ 
   & $\hat{\phi}_{Tertiary}$ & 1.796 *** & 1.795 *** & 1.793 *** & 1.849 *** & 1.851 *** & 1.852 *** & 1.846 *** & 1.846 *** & 1.845 *** \\ 
   &  & (0.233) & (0.232) & (0.232) & (0.249) & (0.252) & (0.251) & (0.238) & (0.237) & (0.236) \\ 
   \bottomrule

    \end{tabularx}
    \begin{minipage}{16.5cm}
            \footnotesize \textbf{Notes:} The estimation is made using a panel of seven Mexican regions, with population-weighted climate, and macroeconomic data from the first quarter of 2000 to the fourth quarter of 2024. The estimated long-run effects, $\hat\theta$, are calculated from the short-run OLS estimates of \autoref{model:ardl}, that is, $\hat{\theta} = \hat{\phi}^{-1} \sum_{l=0}^{p}\hat{\bm{\beta}}_{l}$, where~$\hat{\phi} = 1 - \sum_{l=0}^{p}\hat{\varphi}_{l}$. Cross-sectional and serial correlation in the data were addressed by using  \cite{DriscollKraay} standard errors in the estimations of the short-run parameters, $\hat{\beta}$ and $\hat{\varphi}$. The standard errors associated to the long-run estimations of $\theta$ and $\phi$ are computed with the Delta Method, as suggested by \cite{pesaran2015time}, and are shown in parentheses. Asterisks indicate statistical significance at 1\%~(***), 5\%~(**), and 10\%~(*) levels. Temperature (precipitation) is measured in degrees Celsius (millimeters). The letter $m$ stands for the years used for the construction of the climate norm.
    \end{minipage}
\end{table}

\begin{table}[!htb]
    \centering
    \caption{\small Long-run effects of absolute weather anomalies on GDP per capita by economic sectors.}
    \label{tab:sectors_ARDL_abs}
    \vspace{0.15cm}
    \scriptsize 
    \begin{tabularx}{\textwidth}{mmmmmmmmmmmm}
         \toprule
         & & \multicolumn{3}{c}{Specification 1} & \multicolumn{3}{c}{Specification 2} & \multicolumn{3}{c}{Specification 3} \\
         \cmidrule(lr){3-5} \cmidrule(lr){6-8} \cmidrule(lr){9-11}
          &  & m = 20 & m = 30 & m = 40 & m = 20 & m = 30 & m = 40 & m = 20 & m = 30 & m = 40 \\ 
  \midrule
\multirow{6}{4em}{Primary} & $\hat{\theta}_{\Delta|\Tilde{T}_{it}(m)|}$ & -2.064 *** & -3.084 *** & -4.063 *** & -1.910 *** & -2.831 *** & -3.702 *** &  &  &  \\ 
   &  & (0.743) & (1.093) & (1.449) & (0.737) & (1.083) & (1.437) &  &  &  \\ 
   & $\hat{\theta}_{\Delta|\Tilde{P}_{it}(m)|}$ & -0.008  & -0.011  & -0.015  &  &  &  & -0.004  & -0.004  & -0.007  \\ 
   &  & (0.012) & (0.018) & (0.024) &  &  &  & (0.016) & (0.023) & (0.030) \\ 
   & $\hat{\phi}_{Primary}$ & 2.721 *** & 2.722 *** & 2.722 *** & 2.723 *** & 2.724 *** & 2.724 *** & 2.671 *** & 2.672 *** & 2.671 *** \\ 
   &  & (0.288) & (0.288) & (0.288) & (0.285) & (0.284) & (0.284) & (0.284) & (0.284) & (0.284) \\ 
   \midrule
\multirow{6}{4em}{Secondary} & $\hat{\theta}_{\Delta|\Tilde{T}_{it}(m)|}$ & -1.676  & -2.535  & -3.303  & -1.938  & -2.959  & -3.854  &  &  &  \\ 
   &  & (1.834) & (2.743) & (3.642) & (1.800) & (2.677) & (3.553) &  &  &  \\ 
   & $\hat{\theta}_{\Delta|\Tilde{P}_{it}(m)|}$ & 0.062 * & 0.089 * & 0.116 * &  &  &  & 0.068  & 0.100  & 0.129  \\ 
   &  & (0.035) & (0.052) & (0.068) &  &  &  & (0.044) & (0.065) & (0.085) \\ 
   & $\hat{\phi}_{Secondary}$ & 1.658 *** & 1.658 *** & 1.655 *** & 1.723 *** & 1.723 *** & 1.720 *** & 1.657 *** & 1.657 *** & 1.657 *** \\ 
   &  & (0.329) & (0.329) & (0.328) & (0.344) & (0.345) & (0.344) & (0.368) & (0.368) & (0.368) \\ 
   \midrule
\multirow{6}{4em}{Tertiary} & $\hat{\theta}_{\Delta|\Tilde{T}_{it}(m)|}$ & -0.789  & -1.212  & -1.628  & -0.743  & -1.156  & -1.539  &  &  &  \\ 
   &  & (1.031) & (1.543) & (2.048) & (1.013) & (1.482) & (1.950) &  &  &  \\ 
   & $\hat{\theta}_{\Delta|\Tilde{P}_{it}(m)|}$ & 0.028  & 0.040  & 0.052  &  &  &  & 0.034  & 0.050  & 0.065  \\ 
   &  & (0.023) & (0.034) & (0.044) &  &  &  & (0.027) & (0.040) & (0.052) \\ 
   & $\hat{\phi}_{Tertiary}$ & 1.845 *** & 1.846 *** & 1.843 *** & 1.906 *** & 1.904 *** & 1.901 *** & 1.838 *** & 1.837 *** & 1.837 *** \\ 
   &  & (0.251) & (0.250) & (0.247) & (0.260) & (0.261) & (0.259) & (0.268) & (0.266) & (0.265) \\ 
   \bottomrule

    \end{tabularx}
    \begin{minipage}{16.5cm}
            \scriptsize \textbf{Notes:} The estimation is made using a panel of seven Mexican regions, with population-weighted climate, and macroeconomic data from the first quarter of 2000 to the fourth quarter of 2024. The estimated long-run effects, $\hat\theta$, are calculated from the short-run OLS estimates of \autoref{model:ardl}, that is, $\hat{\theta} = \hat{\phi}^{-1} \sum_{l=0}^{p}\hat{\bm{\beta}}_{l}$, where~$\hat{\phi} = 1 - \sum_{l=0}^{p}\hat{\varphi}_{l}$. Note that, in this case, $\tilde{\vx}_{i,t}(m) = \left[ |\Tilde{T}_{i,t}(m)|,|\Tilde{P}_{i,t}(m)|\right]$ in \autoref{model:ardl}. Cross-sectional and serial correlation in the data were addressed by using  \cite{DriscollKraay} standard errors in the estimations of the short-run parameters, $\beta$ and $\varphi$. The standard errors associated to the long-run estimations of $\theta$ and $\phi$ are computed with the Delta Method, as suggested by \cite{pesaran2015time}, and are shown in parentheses. Asterisks indicate statistical significance at 1\%~(***), 5\%~(**), and 10\%~(*) levels. Temperature (precipitation) is measured in degrees Celsius (millimeters). The letter $m$ stands for the years used for the construction of the climate norm.
    \end{minipage}
\end{table}

\begin{table}[htb]
    \centering
    \caption{\small Long-run effects of weather anomalies on GDP per capita.}
    \label{tab:rest_all_ardl_Noabs}
    \vspace{0.15cm} 
    \scriptsize 
    \begin{tabularx}{\textwidth}{mmmmmmmmmmm}
        \toprule
        & \multicolumn{3}{c}{Specification 1} & \multicolumn{3}{c}{Specification 2} & \multicolumn{3}{c}{Specification 3} \\
         \cmidrule(lr){2-4} \cmidrule(lr){5-7} \cmidrule(lr){8-10}
         & m = 20 & m = 30 & m = 40 & m = 20 & m = 30 & m = 40 & m = 20 & m = 30 & m = 40 \\ 
  \midrule
$\hat{\theta}_{\Delta\Tilde{T}_{it}(m)}$ & -0.796  & -1.052  & -1.391  & -0.743  & -1.015  & -1.347  &  &  &  \\ 
   & (1.090) & (1.530) & (2.025) & (0.732) & (1.029) & (1.349) &  &  &  \\ 
  $\hat{\theta}_{\Delta\Tilde{P}_{it}(m)}$ & -0.009  & -0.009  & -0.014  &  &  &  & 0.001  & 0.004  & 0.004  \\ 
   & (0.046) & (0.066) & (0.086) &  &  &  & (0.026) & (0.037) & (0.048) \\ 
  $\hat{\phi}$ & 1.807 *** & 1.807 *** & 1.806 *** & 1.820 *** & 1.821 *** & 1.820 *** & 1.842 *** & 1.841 *** & 1.842 *** \\ 
   & (0.347) & (0.346) & (0.346) & (0.375) & (0.375) & (0.375) & (0.350) & (0.350) & (0.349) \\ 
   \bottomrule

    \end{tabularx}
    \begin{minipage}{16.5cm}
        \footnotesize \textbf{Notes:} The estimation is made using a panel of seven Mexican regions, with population-weighted climate, and macroeconomic data from the first quarter of 2000 to the fourth quarter of 2024. The estimated long-run effects, $\hat\theta$, are calculated from the short-run OLS estimates of \autoref{model:ardl}, that is, $\hat{\theta} = \hat{\phi}^{-1} \sum_{l=0}^{p}\hat{\bm{\beta}}_{l}$, where~$\hat{\phi} = 1 - \sum_{l=0}^{p}\hat{\varphi}_{l}$. Note that, in this case, $\tilde{\vx}_{i,t}(m) = \left[ \Tilde{T}_{i,t}(m),\Tilde{P}_{i,t}(m)\right]$ in \autoref{model:ardl}. Cross-sectional and serial correlation in the data were addressed by using  \cite{DriscollKraay} standard errors in the estimations of the short-run parameters, $\beta$ and $\varphi$. The standard errors associated to the long-run estimations of $\theta$ and $\phi$ are computed with the Delta Method, as suggested by \cite{pesaran2015time}, and are shown in parentheses. Asterisks indicate statistical significance at 1\%~(***), 5\%~(**), and 10\%~(*) levels. Temperature (precipitation) is measured in degrees Celsius (millimeters). The letter $m$ stands for the years used for the construction of the climate norm.
    \end{minipage}
\end{table}

\begin{table}[!htb]
    \centering
    \caption{\small Long-run effects of weather anomalies on GDP per capita by economic sectors.}
    \label{tab:sectors_ARDL_Noabs}
    \vspace{0.15cm}
    \scriptsize 
    \begin{tabularx}{\textwidth}{mmmmmmmmmmmm}
         \toprule
         & & \multicolumn{3}{c}{Specification 1} & \multicolumn{3}{c}{Specification 2} & \multicolumn{3}{c}{Specification 3} \\
         \cmidrule(lr){3-5} \cmidrule(lr){6-8} \cmidrule(lr){9-11}
          &  & m = 20 & m = 30 & m = 40 & m = 20 & m = 30 & m = 40 & m = 20 & m = 30 & m = 40 \\ 
  \midrule
\multirow{6}{4em}{Primary} & $\hat{\theta}_{\Delta\Tilde{T}_{it}(m)}$ & 0.466  & 0.714  & 0.925  & 0.323  & 0.503  & 0.666  &  &  &  \\ 
   &  & (0.553) & (0.823) & (1.098) & (0.533) & (0.787) & (1.050) &  &  &  \\ 
   & $\hat{\theta}_{\Delta\Tilde{P}_{it}(m)}$ & -0.000  & -0.000  & -0.002  &  &  &  & 0.003  & 0.003  & 0.003  \\ 
   &  & (0.016) & (0.024) & (0.032) &  &  &  & (0.015) & (0.022) & (0.029) \\ 
   & $\hat{\phi}_{Primary}$ & 2.662 *** & 2.662 *** & 2.661 *** & 2.670 *** & 2.671 *** & 2.671 *** & 2.671 *** & 2.670 *** & 2.670 *** \\ 
   &  & (0.280) & (0.280) & (0.280) & (0.279) & (0.279) & (0.279) & (0.280) & (0.280) & (0.281) \\ 
   \midrule
\multirow{6}{4em}{Secondary} & $\hat{\theta}_{\Delta\Tilde{T}_{it}(m)}$ & -0.753  & -0.935  & -1.228  & -0.731  & -0.951  & -1.259  &  &  &  \\ 
   &  & (1.374) & (1.951) & (2.581) & (0.965) & (1.388) & (1.821) &  &  &  \\ 
   & $\hat{\theta}_{\Delta\Tilde{P}_{it}(m)}$ & 0.000  & 0.006  & 0.005  &  &  &  & 0.006  & 0.013  & 0.015  \\ 
   &  & (0.055) & (0.080) & (0.104) &  &  &  & (0.032) & (0.045) & (0.060) \\ 
   & $\hat{\phi}_{Secondary}$ & 1.694 *** & 1.695 *** & 1.695 *** & 1.712 *** & 1.713 *** & 1.713 *** & 1.715 *** & 1.715 *** & 1.715 *** \\ 
   &  & (0.361) & (0.361) & (0.360) & (0.380) & (0.380) & (0.380) & (0.370) & (0.370) & (0.370) \\ 
   \midrule
\multirow{6}{4em}{Tertiary} & $\hat{\theta}_{\Delta\Tilde{T}_{it}(m)}$ & -0.800  & -1.081  & -1.428  & -0.748  & -1.047  & -1.379  &  &  &  \\ 
   &  & (1.001) & (1.399) & (1.850) & (0.683) & (0.955) & (1.251) &  &  &  \\ 
   & $\hat{\theta}_{\Delta\Tilde{P}_{it}(m)}$ & -0.016  & -0.020  & -0.028  &  &  &  & -0.003  & -0.002  & -0.004  \\ 
   &  & (0.044) & (0.063) & (0.082) &  &  &  & (0.026) & (0.037) & (0.048) \\ 
   & $\hat{\phi}_{Tertiary}$ & 1.865 *** & 1.864 *** & 1.863 *** & 1.862 *** & 1.863 *** & 1.862 *** & 1.906 *** & 1.904 *** & 1.905 *** \\ 
   &  & (0.253) & (0.252) & (0.251) & (0.273) & (0.274) & (0.274) & (0.250) & (0.248) & (0.248) \\ 
   \bottomrule

    \end{tabularx}
    \begin{minipage}{16.5cm}
            \scriptsize \textbf{Notes:} The estimation is made using a panel of seven Mexican regions, with population-weighted climate, and macroeconomic data from the first quarter of 2000 to the fourth quarter of 2024. The estimated long-run effects, $\hat\theta$, are calculated from the short-run OLS estimates of \autoref{model:ardl}, that is, $\hat{\theta} = \hat{\phi}^{-1} \sum_{l=0}^{p}\hat{\bm{\beta}}_{l}$, where~$\hat{\phi} = 1 - \sum_{l=0}^{p}\hat{\varphi}_{l}$. Note that, in this case, $\tilde{\vx}_{i,t}(m) = \left[ \Tilde{T}_{i,t}(m),\Tilde{P}_{i,t}(m)\right]$ in \autoref{model:ardl}. Cross-sectional and serial correlation in the data were addressed by using  \cite{DriscollKraay} standard errors in the estimations of the short-run parameters, $\beta$ and $\varphi$. The standard errors associated to the long-run estimations of $\theta$ and $\phi$ are computed with the Delta Method, as suggested by \cite{pesaran2015time}, and are shown in parentheses. Asterisks indicate statistical significance at 1\%~(***), 5\%~(**), and 10\%~(*) levels. Temperature (precipitation) is measured in degrees Celsius (millimeters). The letter $m$ stands for the years used for the construction of the climate norm.
    \end{minipage}
\end{table}

\begin{table}[!htb]
    \centering
    \caption{\small{Response of CPI inflation to an unit impulse to positive and negative weather anomalies.}}
    \label{tab:rest_inpc_deviationsLP}
    \vspace{0.15cm}
    \footnotesize  
    \begin{tabular}{llccccccccc}
    \toprule
          &  & $h=0$ & $h=1$ & $h=2$ & $h=3$ & $h=4$ & $h=5$ & $h=6$ & $h=7$ & $h=8$ \\ 
  \midrule
\multirow{12}{3em}{$\Tilde{T}_{it}^{+}(m)$} & All items & 0.24  & -0.47  & -1.08  & -1.23  & -0.82  & 0.44  & -0.38  & -1.77  & -1.11  \\ 
   &  & (1.520) & (2.606) & (3.219) & (4.094) & (4.720) & (4.963) & (5.467) & (6.078) & (6.605) \\ 
   & Food & -0.60  & -1.79  & -1.93  & 0.05  & -0.18  & -1.21  & -1.69  & 1.30  & 2.52  \\ 
   &  & (1.903) & (3.973) & (5.484) & (7.142) & (8.240) & (9.362) & (10.369) & (11.607) & (12.309) \\ 
   & Non food & -0.10  & 0.12  & -0.22  & -0.37  & -0.43  & -0.47  & -0.75  & 0.29  & 0.20  \\ 
   &  & (0.947) & (1.880) & (2.810) & (3.814) & (4.689) & (5.588) & (6.249) & (7.008) & (7.714) \\ 
   & Services & -0.49  & -1.27  & -1.99  & -0.91  & -1.27  & -1.27  & -2.21  & -1.81  & -2.40  \\ 
   &  & (0.966) & (1.645) & (2.235) & (2.800) & (3.315) & (3.702) & (4.204) & (5.004) & (5.528) \\ 
   & Agriculture & 6.76  & 2.61  & -1.14  & -3.10  & 3.92  & 7.77  & -2.44  & -5.19  & 1.56  \\ 
   &  & (8.184) & (11.484) & (11.790) & (13.078) & (12.383) & (13.187) & (11.918) & (13.773) & (12.529) \\ 
   & Energy & -6.56  & -14.29  & -11.68  & -3.34  & -5.15  & -15.30  & -12.52  & 0.19  & 5.21  \\ 
   &  & (9.565) & (12.840) & (14.201) & (14.793) & (17.777) & (14.333) & (16.489) & (15.710) & (15.607) \\ 
   \midrule
\multirow{12}{3em}{$\Tilde{T}_{it}^{-}(m)$} & All items & -0.03  & 1.19  & 2.34  & 1.72  & 1.50  & 0.63  & 0.57  & 0.66  & 1.46  \\ 
   &  & (2.077) & (3.713) & (4.503) & (6.063) & (6.965) & (7.695) & (8.770) & (9.620) & (10.228) \\ 
   & Food & 0.67  & 3.27  & 2.49  & 0.82  & 0.69  & 2.02  & 3.01  & -1.13  & -3.12  \\ 
   &  & (2.854) & (6.232) & (8.950) & (11.771) & (13.549) & (15.388) & (16.650) & (18.189) & (18.858) \\ 
   & Non food & 0.18  & -0.11  & -1.01  & -0.88  & -0.52  & -0.26  & -0.58  & -2.42  & -2.91  \\ 
   &  & (1.297) & (2.838) & (4.150) & (5.609) & (7.069) & (8.423) & (9.452) & (10.254) & (11.096) \\ 
   & Services & 0.55  & 2.32  & 3.04  & 2.35  & 2.90  & 3.41  & 2.72  & 3.02  & 4.44  \\ 
   &  & (1.723) & (2.770) & (3.777) & (4.665) & (5.795) & (6.319) & (7.242) & (8.596) & (9.717) \\ 
   & Agriculture & -1.75  & 1.35  & 7.18  & 4.59  & -10.46  & -6.57  & 12.67  & -0.15  & 0.52  \\ 
   &  & (12.088) & (17.766) & (19.979) & (20.683) & (20.177) & (22.006) & (20.600) & (22.963) & (19.616) \\ 
   & Energy & 6.89  & 14.20  & 12.86  & -0.04  & 15.82  & 17.94  & 3.56  & -7.71  & -8.17  \\ 
   &  & (10.748) & (14.181) & (17.144) & (22.688) & (21.870) & (19.278) & (23.384) & (22.677) & (19.825) \\ 
   \midrule
\multirow{12}{3em}{$\Tilde{P}_{it}^{+}(m)$} & All items & 0.03  & -0.01  & 0.01  & 0.01  & 0.03  & 0.03  & 0.01  & 0.03  & 0.02  \\ 
   &  & (0.033) & (0.053) & (0.057) & (0.081) & (0.091) & (0.094) & (0.110) & (0.122) & (0.132) \\ 
   & Food & -0.02  & -0.06  & -0.03  & -0.04  & -0.02  & -0.03  & -0.01  & 0.01  & 0.02  \\ 
   &  & (0.040) & (0.089) & (0.115) & (0.147) & (0.174) & (0.203) & (0.227) & (0.248) & (0.263) \\ 
   & Non food & 0.02  & 0.01  & 0.03  & 0.02  & 0.04  & 0.01  & 0.04  & 0.02  & 0.04  \\ 
   &  & (0.027) & (0.040) & (0.065) & (0.087) & (0.104) & (0.123) & (0.139) & (0.151) & (0.158) \\ 
   & Services & 0.01  & -0.03  & 0.01  & 0.01  & 0.01  & -0.02  & -0.01  & -0.01  & -0.01  \\ 
   &  & (0.014) & (0.031) & (0.043) & (0.054) & (0.065) & (0.072) & (0.086) & (0.098) & (0.112) \\ 
   & Agriculture & -0.01  & -0.09  & -0.14  & -0.06  & 0.12  & -0.15  & -0.17  & -0.09  & 0.03  \\ 
   &  & (0.182) & (0.199) & (0.240) & (0.270) & (0.262) & (0.239) & (0.269) & (0.306) & (0.246) \\ 
   & Energy & 0.07  & -0.29  & 0.21  & 0.06  & 0.09  & -0.14  & 0.12  & 0.19  & 0.16  \\ 
   &  & (0.174) & (0.298) & (0.312) & (0.343) & (0.301) & (0.361) & (0.303) & (0.324) & (0.332) \\ 
   \midrule
\multirow{12}{3em}{$\Tilde{P}_{it}^{-}(m)$} & All items & -0.01  & -0.01  & 0.04  & 0.03  & -0.01  & -0.02  & 0.06  & 0.07  & 0.03  \\ 
   &  & (0.059) & (0.090) & (0.118) & (0.151) & (0.177) & (0.195) & (0.210) & (0.226) & (0.232) \\ 
   & Food & 0.03  & 0.03  & 0.05  & 0.01  & 0.02  & 0.10  & 0.08  & -0.01  & 0.07  \\ 
   &  & (0.106) & (0.172) & (0.238) & (0.304) & (0.355) & (0.398) & (0.429) & (0.465) & (0.477) \\ 
   & Non food & 0.01  & -0.01  & 0.01  & -0.01  & -0.02  & -0.02  & 0.02  & -0.03  & 0.04  \\ 
   &  & (0.035) & (0.075) & (0.106) & (0.151) & (0.187) & (0.219) & (0.241) & (0.269) & (0.283) \\ 
   & Services & 0.04  & 0.08  & 0.08  & 0.03  & 0.07  & 0.09  & 0.12  & 0.10  & 0.08  \\ 
   &  & (0.034) & (0.062) & (0.086) & (0.106) & (0.131) & (0.149) & (0.170) & (0.198) & (0.212) \\ 
   & Agriculture & -0.09  & 0.01  & 0.15  & -0.01  & -0.26  & 0.03  & 0.23  & 0.24  & -0.07  \\ 
   &  & (0.336) & (0.375) & (0.410) & (0.407) & (0.447) & (0.421) & (0.441) & (0.462) & (0.415) \\ 
   & Energy & 0.44  & 0.61  & 0.51  & -0.01  & 0.44  & 0.52  & 0.49  & -0.42  & 0.04  \\ 
   &  & (0.315) & (0.572) & (0.460) & (0.683) & (0.624) & (0.573) & (0.593) & (0.617) & (0.553) \\ 
   \bottomrule

    \end{tabular}
    \begin{minipage}{16.5cm}
        \footnotesize \textbf{Notes:} Cumulative inflation impulse-responses to population-weighted temperature and precipitation deviations shocks in a panel of seven regions of Mexico. $h$ refers to the horizon in quarters. \cite{DriscollKraay} standard errors are reported in parentheses, and asterisks indicate the statistical significance of the estimates at 1\%~(***), 5\%~(**), and 10\%~(*) levels. Temperature (precipitation) is measured in degrees Celsius (millimeters). The fitted model is similar to \cite{faccia2021feeling} but using deviations from the 30-year temperature/precipitation norm (see~\autoref{sec:method} for variable description). 
    \end{minipage}
\end{table}

\begin{figure}[t]
    \centering
    \subfloat[(a) Total]{
      \includegraphics[width=0.42\textwidth]{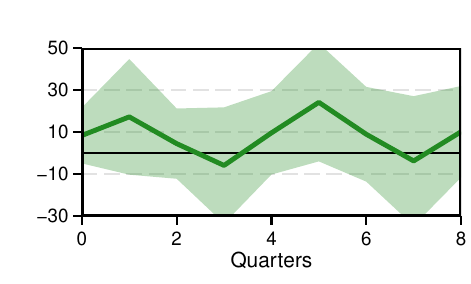}
    }\hfill
    \subfloat[(e) Total]{
      \includegraphics[width=0.42\textwidth]{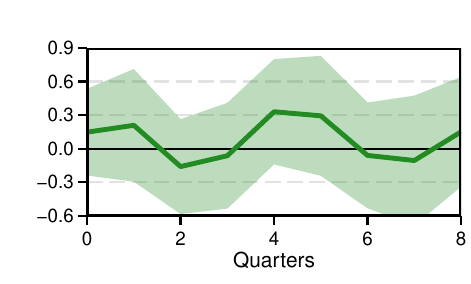}
    }
    \hfil 
    \subfloat[(b) Primary sector]{
      \includegraphics[width=0.42\textwidth]{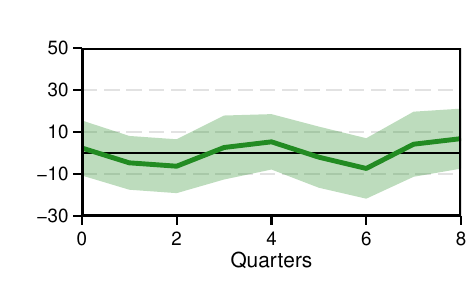}
    }\hfill
    \subfloat[(f) Primary sector]{
      \includegraphics[width=0.42\textwidth]{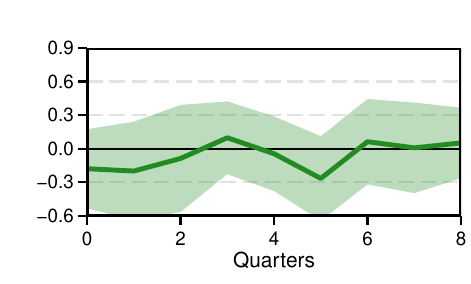}
    }
    \hfil 
    \subfloat[(c) Secondary sector]{
      \includegraphics[width=0.42\textwidth]{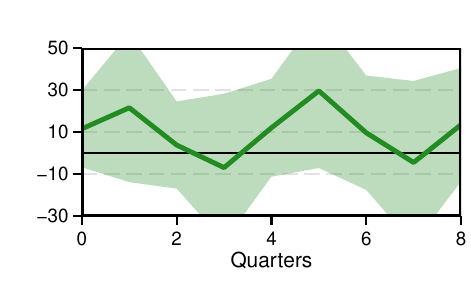}
    }\hfill
    \subfloat[(g) Secondary sector]{
      \includegraphics[width=0.42\textwidth]{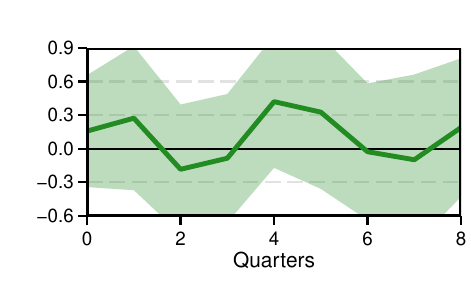}
    }
    \hfil 
    \subfloat[(d) Tertiary sector]{
      \includegraphics[width=0.42\textwidth]{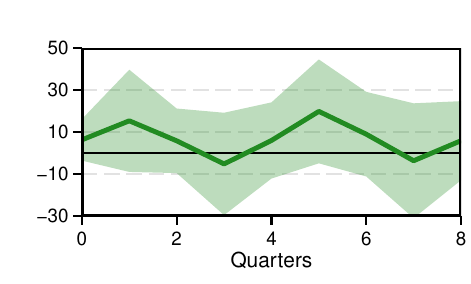}
    }\hfill
    \subfloat[(h) Tertiary sector]{
      \includegraphics[width=0.42\textwidth]{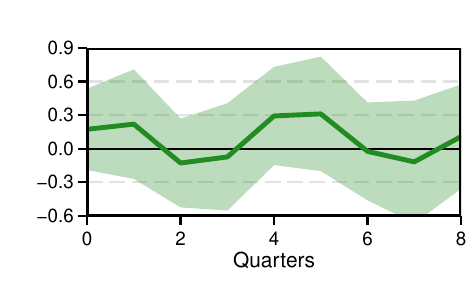}
    }
    \begin{center}
        \caption{Effect of weather anomalies on cumulative GDP per capita by sectors.} \label{fig:LP_GDP-dev}
        \vspace{-0.15cm}
        \caption*{\footnotesize{\textbf{Notes:} Panels (a) to (d) represent the IRs after a shock on temperature deviations, while panels (e) to (h) are the IRs after a shock on precipitation deviations. Cumulative impulse-responses to population-weighted climate deviations from its historical norm in seven regions of Mexico. The shaded area represents 90\% confidence intervals using \cite{DriscollKraay} standard errors.}}
    \end{center}
\end{figure}

\begin{table}[!htb]
    \centering
    \caption{\small{Response of real GDP per capita to a unit impulse to positive and negative weather anomalies.}}
    \label{tab:rest_gdp_deviationsLP}
    \vspace{0.15cm}
    \footnotesize  
    \begin{tabular}{llccccccccc}
    \toprule
          &  & $h=0$ & $h=1$ & $h=2$ & $h=3$ & $h=4$ & $h=5$ & $h=6$ & $h=7$ & $h=8$ \\ 
  \midrule
\multirow{8}{3em}{$\Tilde{T}_{it}^{+}(m)$} & Total & 13.89  & 30.95  & 6.52  & -8.63  & 15.43  & 42.03 * & 15.72  & -5.03  & 14.83  \\ 
   &  & (10.053) & (24.551) & (13.105) & (23.144) & (15.720) & (23.393) & (18.854) & (27.445) & (18.556) \\ 
   & Primary & 2.05  & -4.92  & -13.30  & 9.28  & 2.70  & -2.87  & -13.35  & 8.47  & 5.94  \\ 
   &  & (10.186) & (9.099) & (9.940) & (10.936) & (9.993) & (11.726) & (11.060) & (12.740) & (10.948) \\ 
   & Secondary & 19.06  & 40.64  & 4.80  & -10.48  & 20.35  & 52.51 * & 18.04  & -5.63  & 20.82  \\ 
   &  & (14.468) & (32.748) & (16.241) & (30.826) & (19.160) & (31.419) & (23.381) & (36.368) & (23.767) \\ 
   & Tertiary & 10.83  & 27.56  & 9.13  & -7.65  & 9.61  & 34.98 * & 15.51  & -4.69  & 7.72  \\ 
   &  & (7.387) & (21.718) & (12.649) & (20.477) & (14.350) & (20.649) & (17.035) & (24.246) & (16.048) \\ 
   \midrule
\multirow{8}{3em}{$\Tilde{T}_{it}^{-}(m)$} & Total & -13.54  & -26.06  & -8.68  & 12.54  & -16.99  & -38.82  & -13.54  & 9.35  & -20.12  \\ 
   &  & (16.264) & (29.586) & (19.809) & (33.289) & (24.883) & (35.111) & (26.202) & (35.174) & (24.624) \\ 
   & Primary & -7.43  & 15.09  & 2.40  & 5.87  & -19.44  & 4.69  & 8.68  & -4.07  & -20.42  \\ 
   &  & (16.668) & (18.095) & (15.904) & (22.540) & (18.090) & (18.474) & (20.292) & (19.364) & (19.652) \\ 
   & Secondary & -19.30  & -31.20  & -8.75  & 13.74  & -20.04  & -46.39  & -13.12  & 12.18  & -25.23  \\ 
   &  & (22.181) & (39.621) & (26.008) & (43.264) & (29.191) & (48.331) & (32.454) & (45.046) & (31.150) \\ 
   & Tertiary & -9.23  & -23.22  & -10.33  & 11.05  & -10.38  & -31.70  & -14.18  & 9.78  & -13.14  \\ 
   &  & (12.186) & (25.163) & (16.720) & (28.411) & (22.498) & (30.090) & (22.574) & (30.411) & (20.821) \\ 
   \midrule
\multirow{8}{3em}{$\Tilde{P}_{it}^{+}(m)$} & Total & -0.01  & 0.22  & -0.28  & -0.01  & 0.21  & 0.31  & -0.20  & -0.03  & 0.09  \\ 
   &  & (0.165) & (0.317) & (0.286) & (0.200) & (0.254) & (0.360) & (0.291) & (0.265) & (0.312) \\ 
   & Primary & -0.18  & -0.20  & -0.11  & 0.13  & -0.05  & -0.29  & 0.08  & -0.02  & 0.06  \\ 
   &  & (0.249) & (0.271) & (0.343) & (0.213) & (0.194) & (0.247) & (0.262) & (0.270) & (0.205) \\ 
   & Secondary & 0.01  & 0.31  & -0.29  & -0.04  & 0.34  & 0.36  & -0.14  & -0.02  & 0.15  \\ 
   &  & (0.220) & (0.409) & (0.400) & (0.272) & (0.312) & (0.464) & (0.392) & (0.349) & (0.386) \\ 
   & Tertiary & 0.02  & 0.21  & -0.24  & -0.03  & 0.18  & 0.31  & -0.15  & -0.06  & 0.06  \\ 
   &  & (0.153) & (0.300) & (0.261) & (0.204) & (0.233) & (0.333) & (0.264) & (0.251) & (0.295) \\ 
   \midrule
\multirow{8}{3em}{$\Tilde{P}_{it}^{-}(m)$} & Total & -0.83  & -0.45  & 0.04  & 0.29  & -1.12  & -0.64  & -0.26  & 0.47  & -0.53  \\ 
   &  & (0.663) & (0.486) & (0.403) & (0.804) & (0.754) & (0.557) & (0.609) & (0.969) & (0.589) \\ 
   & Primary & 0.42  & 0.46  & 0.15  & -0.12  & 0.11  & 0.56  & -0.10  & -0.10  & -0.08  \\ 
   &  & (0.381) & (0.452) & (0.382) & (0.351) & (0.438) & (0.430) & (0.352) & (0.477) & (0.376) \\ 
   & Secondary & -0.82  & -0.52  & 0.12  & 0.34  & -1.20  & -0.65  & -0.26  & 0.45  & -0.56  \\ 
   &  & (0.864) & (0.619) & (0.502) & (0.935) & (0.975) & (0.717) & (0.722) & (1.265) & (0.763) \\ 
   & Tertiary & -0.88  & -0.53  & -0.03  & 0.30  & -1.02  & -0.73  & -0.30  & 0.45  & -0.39  \\ 
   &  & (0.645) & (0.498) & (0.399) & (0.806) & (0.713) & (0.551) & (0.589) & (0.919) & (0.569) \\ 
   \bottomrule

    \end{tabular}
    \begin{minipage}{16.5cm}
        \footnotesize \textbf{Notes:} Cumulative inflation impulse-responses to population-weighted temperature and precipitation deviations shocks in a panel of seven regions of Mexico. $h$ refers to the horizon in quarters. \cite{DriscollKraay} standard errors are reported in parentheses, and asterisks indicate the statistical significance of the estimates at 1\%~(***), 5\%~(**), and 10\%~(*) levels. Temperature (precipitation) is measured in degrees Celsius (millimeters). The fitted model is similar to \cite{faccia2021feeling} but using deviations from the 30-year temperature/precipitation norm (see~\autoref{sec:method} for variable description). 
    \end{minipage}
\end{table}


\end{document}